\documentclass[lettersize,journal]{IEEEtran}
\usepackage{amsmath,amsfonts}
\usepackage{algorithmic}
\usepackage{algorithm}
\usepackage{array}

\usepackage{textcomp}
\usepackage{stfloats}
\usepackage{url}
\usepackage{verbatim}
\usepackage{graphicx}
\usepackage{cite}
\usepackage{amsmath}
\usepackage{subfigure}
\usepackage{booktabs}
\usepackage{multirow}
\begin{document}

\title{Channel Semantic Characterization for Integrated Sensing and Communication Scenarios: From Measurements to Modeling}

\author{Zhengyu~Zhang,~\IEEEmembership{Student Member, IEEE,}
        Ruisi~He,~\IEEEmembership{Senior Member, IEEE,}
        Bo~Ai,~\IEEEmembership{Fellow, IEEE,}
        Mi~Yang,~\IEEEmembership{Member, IEEE,}
        Xuejian~Zhang,~\IEEEmembership{Student Member, IEEE,}
        Ziyi~Qi,
        Zhangdui~Zhong,~\IEEEmembership{Fellow, IEEE,}

\thanks{
Zhengyu Zhang, Ruisi He, Bo Ai, Mi Yang, Xuejian Zhang, Ziyi Qi and Zhangdui Zhong are with the School of Electronics and Information Engineering, Beijing Jiaotong University, Beijing 100044, China. (e-mail:ruisi.he@bjtu.edu.cn)}

\markboth{Journal of \LaTeX\ Class Files,~Vol.~14, No.~8, August~2021}%
{Shell \MakeLowercase{\textit{et al.}}: A Sample Article Using IEEEtran.cls for IEEE Journals}

\IEEEpubid{0000--0000/00\$00.00~\copyright~2021 IEEE}}

\maketitle

\begin{abstract}
With the advancement of sixth-generation (6G) wireless communication systems, integrated sensing and communication (ISAC) is crucial for perceiving and interacting with the environment via electromagnetic propagation, termed channel semantics, to support tasks like decision-making. However, channel models focusing on physical characteristics face challenges in representing semantics embedded in the channel, thereby limiting the evaluation of ISAC systems. To tackle this, we present a novel framework for channel modeling from the conceptual event perspective. By leveraging a multi-level semantic structure and characterized knowledge libraries, the framework decomposes complex channel characteristics into extensible semantic characterization, thereby better capturing the relationship between environment and channel, and enabling more flexible adjustments of channel models for different events without requiring a complete reset. Specifically, we define channel semantics on three levels: status semantics, behavior semantics, and event semantics, corresponding to channel multipaths, channel time-varying trajectories, and channel topology, respectively. Taking realistic vehicular ISAC scenarios as an example, we perform semantic clustering, characterizing status semantics via multipath statistical distributions, modeling behavior semantics using Markov chains for time variation, and representing event semantics through a co-occurrence matrix. Results show the model accurately generates channels while capturing rich semantic information. Moreover, its generalization supports customized semantics.

\end{abstract}


\section{Introduction}

\subsection{Background and Motivation}
\IEEEPARstart{T}{he} rapid development of mobile communication technology is driving a paradigm shift from fifth-generation (5G) to sixth-generation (6G) wireless communication systems \cite{1}. One of the key innovations in 6G is the integration of sensing and communication functions into shared time / frequency / platform, known as Integrated Sensing and Communication (ISAC)\cite{3}. It enables wireless communication and environmental sensing simultaneously without the need for additional hardware, and improves the native sensing capabilities of wireless devices\cite{4}. Benefiting from the dual functionality of the ISAC, it becomes possible for wireless communication systems to "see the environment", which is essential for emerging applications such as intelligent transportation, the Internet of Things (IoT), smart cities, etc.\cite{5,6,G2}. Furthermore, with advances in artificial intelligence (AI) and computer vision (CV), ISAC systems are expected to evolve from merely 'seeing' to actively 'understanding' the environment, enabling intelligent agents to perceive, predict, and interact with their surroundings\cite{8}. Building on this vision, \textit{semantics}—referring to the interpretation of specific events and their contributions to electromagnetic propagation—has emerged as a novel and promising dimension of information for 6G networks \cite{add1,9}. Assisted by this advancement, ISAC systems will not only utilize electromagnetic propagation for data transmission but also leverage it as a foundation for sensing and understanding the environment, thereby supporting tasks such as decision-making, beamforming, and recognition\cite{10,11,12}.

For ISAC systems, it is critical to accurately understand the interaction between the physical environment and the wireless channel. Specifically, the physical environment shapes the characteristics of the wireless channel through propagation mechanisms such as reflection, scattering, and diffraction. Conversely, these wireless channel characteristics reflect the surrounding environment through multipath components, cluster distributions, and others \cite{13,14,15}. For example, in vehicular communication scenarios, multipaths from surrounding objects, such as vehicles, buildings, and road infrastructure, dynamically reflect rapid changes in the environment\cite{A1,A5}. Furthermore, semantic information—such as the presence, position, and movement of objects—can be extracted from channel and interpreted by ISAC systems, enabling responsive actions such as autonomous driving. This bidirectional interaction between the environment and the channel underscores the importance of semantic understanding in ISAC systems, bridging the gap between low-level physical signals and high-level environmental awareness. As a result, semantics have become an indispensable characteristic of ISAC channels.

However, traditional channel modeling approaches, such as statistical channel models\cite{16} and deterministic channel models\cite{17}, fall short in capturing semantic characterization. They focus on physical properties, such as path loss, multipath fading, and delay spread, but struggle to represent the semantic information embedded in the channel, thereby limiting the performance evaluation of ISAC systems. On the other hand, AI-based channels provide a data-driven approach to channel modeling, enabling automatic fitting of channel characteristics\cite{A2,A3,A6}. Although AI-based channels have made significant progress in terms of accuracy and predictive capabilities, their "black-box" nature makes it difficult to provide semantic interpretations of channels. Therefore, developing a channel semantic model has become a great challenge. Motivated by this emerging focus, this paper investigates channel semantic characterization for ISAC scenarios, and explored the channel semantic modeling approach based on actual measurements.

\subsection{Related work}

Modeling of ISAC channels with semantic characteristics is crucial. However, to the best of the authors’ knowledge, there are limited studies focusing on ISAC channel measurements and modeling specifically for semantic characterization. This subsection summarizes some related investigations on i) Channel measurements for ISAC scenarios, ii) Channel modeling for ISAC scenarios and iii) Characterization for channel semantics, as follows:

\begin{itemize}
\item[1)] 
Measurement is an effective method to characterize wireless channel \cite{19,20,21}. In order to explore special characteristics of sensing channels, there exists some measurement campaigns for ISAC scenarios, which mainly extract channel parameters based on echoes. In \cite{22}, channel measurement campaigns at millimeter-wave frequency bands were conducted in an outdoor ISAC scenario, with a comparison presented between communication and sensing channels. Ref. \cite{23} conducted the sensing channel measurements with an ISAC prototype in the indoor hotspots (InH) and urban micro (UMi) scenarios, in which the sensing pathloss was focused. In \cite{24}, ISAC channel characteristics were analyzed using data collected from two sets of measurements conducted at 28 GHz in an outdoor scenario. Furthermore, a metal board was introduced to study the effect of an interference environment on the sensing channel. Ref. \cite{25} conducted ISAC channel measurements at 10 GHz in an indoor scenario, thoroughly exploring the statistical differences between the two channels. However, these measurement campaigns for ISAC scenarios primarily focus on basic physical channel parameters, limiting to simultaneously capture the semantic information embedded in the channel from the surrounding environment.
\end{itemize}

\begin{itemize}
\item[2)] 
Aimed to support the environmental sensing of ISAC, wireless channel models have increasingly focused on the relationship between radio propagation and physical environments. In \cite{26}, the authors introduced a novel channel model for mapping diffuse and specular scattering, which enabled efficient tracking of individual scatterers over time using an interacting multiple model (IMM) extended Kalman filter and smoother. Ref. \cite{27} proposed a cluster-based ISAC statistical channel modeling using measured data by mapping the multipath clusters to scatterers in the environment, which enabled modeling the correlation and differences between the communication and sensing channels. Ref. \cite{28} presented a hybrid channel modeling method that incorporates statistical and deterministic modeling, and adopted the multi-scattering center theory to model the target's electromagnetic features. In \cite{29}, a stochastic ISAC channel model was proposed to capture the sharing feature, where shared and non-shared clusters by the two channels were defined and superimposed based on the surrounding environments. However, these studies primarily focus on scatterer localization or reconstruction, leading to a lack of semantic characterization under specific event. Consequently, establishing a mapping between the channel models and semantic characteristics remains a significant challenge.
\end{itemize}

\begin{itemize}
\item[3)] 

In the area of semantic communication, semantic-aware joint source and channel coding are implemented to enable effective and accurate information transfer based on a semantic knowledge library\cite{C1,31,G1}. In such case, the channel semantic model is viewed as a pipeline for semantic transmission, where physical signals contain redundant information compared to the intended meaning\cite{33}. However, in ISAC scenarios, to achieve highly efficient environmental perception and understanding, the channel semantic model is expected to accurately reflect the surrounding environment. There are some channel semantic model designed for a specific task or communication system. For example, Ref. \cite{34} defined task-oriented propagation channel semantic model and developed an environment semantics-assisted beam prediction method based on relevant environmental characteristics. Ref. \cite{35} proposed a predictive channel-based semantic communication system tailored for specific sensing scenarios, leveraging semantic information to meet specific application requirements. From the aspects of waveform design and signal processing, Ref. \cite{36} introduced a two-stage AI frame structure to achieve synergistic gains by designing a joint channel semantic reconstruction model. However, these channel semantic models are primarily tailored for specific communication tasks. To support advanced ISAC technology in 6G, further research on channel modeling is required to explore semantic characterization influenced by environmental elements.

\end{itemize}

\subsection{Major Contributions}

As discussed earlier, incorporating semantic characteristics into wireless channel models is crucial for the design and validation of emerging ISAC systems. In order to fill the aforementioned gaps, this paper decomposes complex channel characteristics into composable and extensible high-level semantic characterization, thereby better capturing the relationship between the environment and channel, and enabling more flexible adjustments of channel models for different events without requiring a complete reset. The work introduces an innovative channel semantic modeling approach, in which model parameters are extracted from realistic measurements and a generalized implementation of channel simulation is validated. The main contributions are summarized as follows:

\begin{itemize}
\item[1)]
A new framework of channel semantic model for ISAC scenarios is proposed, which composes of status semantics, behavior semantics and event semantics, corresponding to channel multipaths, channel trajectories and channel topologies respectively.
\end{itemize}

\begin{itemize}
\item[2)]
A vehicular channel measurement system is employed to perform ISAC channel measurements at 28 GHz, and semantic information of the measured environment is extracted using CV algorithms.
\end{itemize}

\begin{itemize}
\item[3)]

Based on realistic measurements, channel semantic is characterized.  After semantic clustering, status semantics are characterized by statistical distributions for multipath number, delays, and power. Behavior semantics use Markov chains to model time-varying multipath characteristics. Event semantics employ a co-occurrence matrix to link behaviors with status, forming a comprehensive semantic representation.

\end{itemize}

\begin{itemize}
\item[4)]
The implementation of channel simulation from the semantic level is presented. The results indicate that the proposed model can generate accurate channels while representing rich semantic information. Additionally, generalization of the model for customized semantics is demonstrated.

\end{itemize}
The remainder of this paper is structured as follows. In Section II, a channel semantic model is proposed for ISAC scenarios. Section III introduces ISAC channel measurement system and measurement campaign. In Section IV, measurement data processing and characterization for different semantics are presented. Section V presents channel semantic model implementation and validation. Finally, conclusion is given in Section VI.

\section{Channel Semantic Models}

\begin{figure}[t]
    \centering
        \includegraphics[width=1\linewidth]{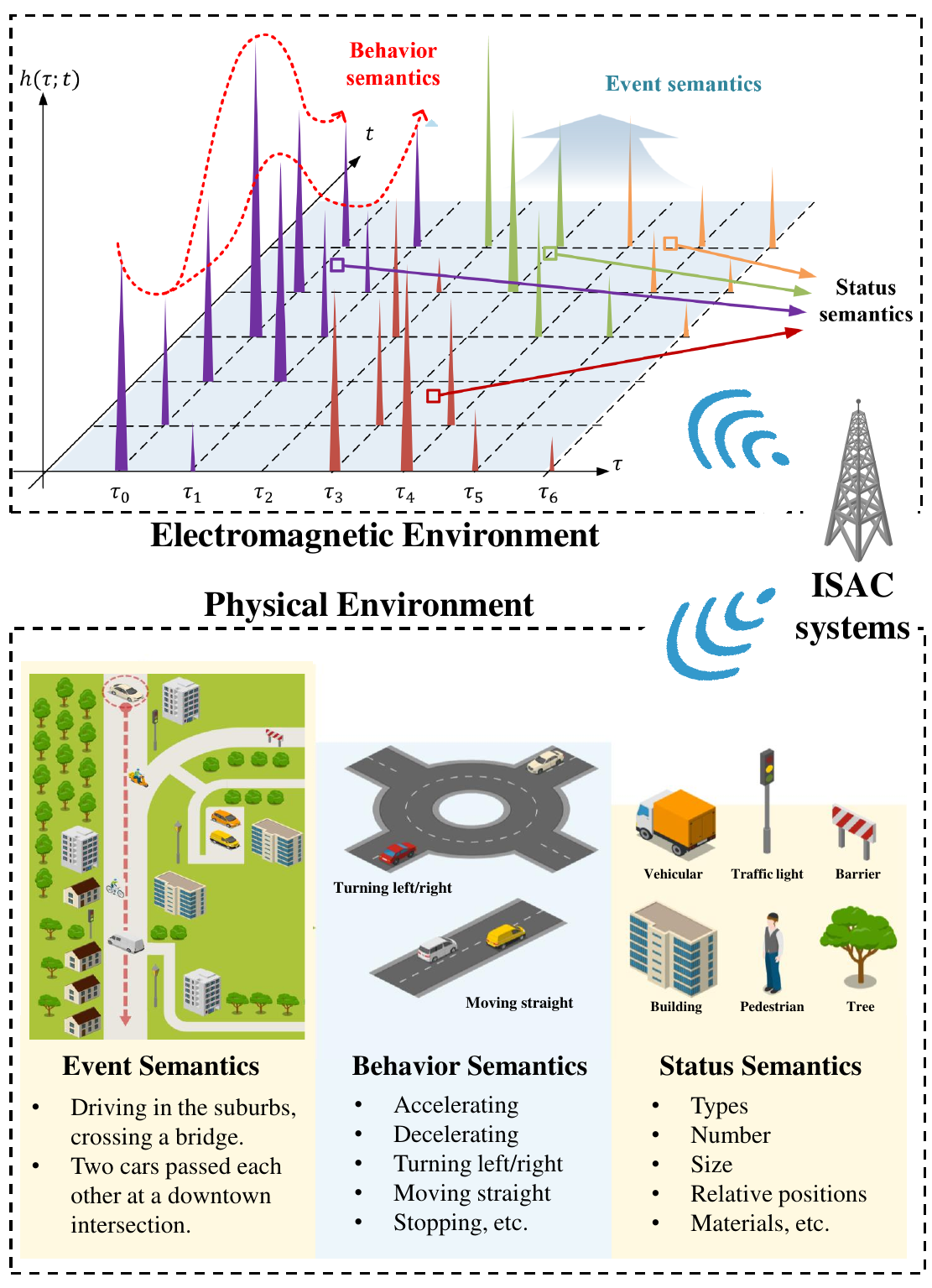}
    \caption{The illustration of channel semantics.}
    \label{fig1}
\end{figure}
\begin{figure}[t]
    \centering
        \includegraphics[width=1\linewidth]{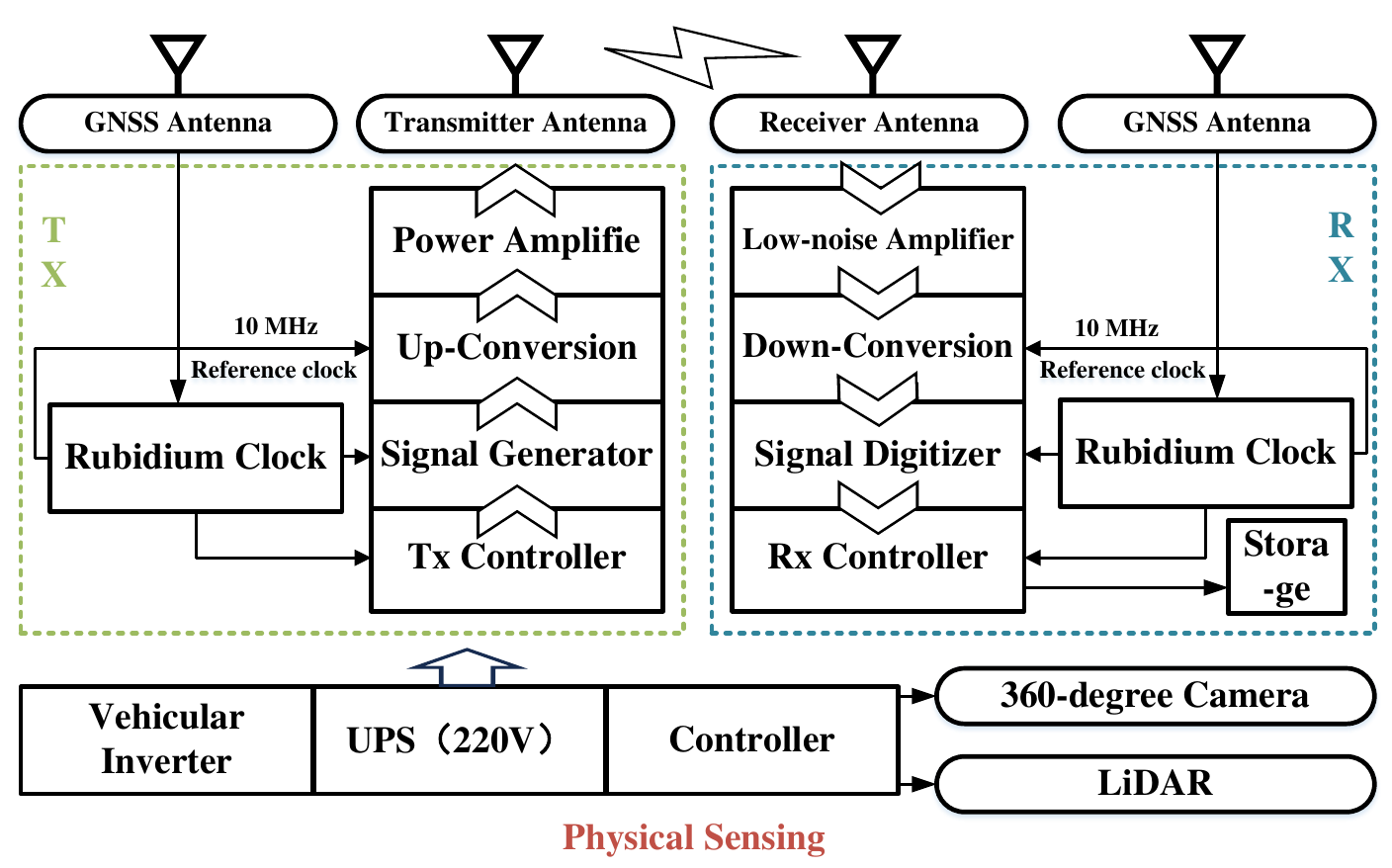}
    \caption{Vehicular channel measurement system for ISAC scenarios.}
    \label{fig1}
\end{figure}

To model ISAC channels with semantic characteristics, we propose a novel channel semantic model. As illustrated in Fig. 1, the model decouples traditional time-varying channel models into multilevel semantic information. Channel semantics are hierarchically defined from low to high levels as status semantics, behavior semantics, and event semantics in the physical environment, corresponding to channel multipath, channel time-varying trajectories, and channel topology in the electromagnetic environment, respectively. These semantics collectively describe the semantic representation of the propagation environment, with lower-level semantics forming the foundation for higher-level semantics, and higher-level semantics integrating multiple lower-level semantics.

To achieve interpretability and scalability, the channel semantic model should:

$\bullet$ represent the implicit correlations between statuses and behaviors within typical event scenarios;

$\bullet$ be able to store historical behaviors for reasoning and prediction;

$\bullet$ contain potential statuses in the environment based on a structured knowledge library.

To formally represent the relationships between events, behaviors, and statuses in the channel semantic model, we define the following sets and their relationships:

The event set \( \mathbf{E} \) represents a set of \( n \) events, where each event \( E_i \) can be correlated, independent, simultaneous, or sequential. These events represent macroscopic behaviors relative to the current observation view and describe the topological relationships between behaviors and statuses. This is formally represented as:
\[
\mathbf{E} = \left\{ E_1, E_2, \dots, E_n \right\} \quad \text{where} \quad E_i = f_i(\mathbf{B}, \mathbf{S}) \
\]
where $f_i$ represents the topological relationship mapping between statuses and behaviors within events. For example, in Fig. 1, the entire process of two cars passing each other at a downtown intersection can be considered one event. Similarly, driving in the suburbs and crossing a bridge can also be regarded as another event. An $\mathit{Event}$ can also refer to processes that span larger temporal and spatial scales. Events involving multiple behaviors and status can represent a semantic aggregation characterized by interrelated relationships.

The behavior set \( \mathbf{B} \) represents a set of \( m \) behaviors, where each behavior \( B_i \) occurs within a specific event, corresponds to a specific statu, and may jointly influence the event. These behaviors describe actions varying continuously over time and occur along the channel trajectory within a range of delay and time. This is formally represented as:
\[
\mathbf{B} = \left\{ B_1, B_2, \dots, B_m \right\} \quad \text{where} \quad  B_i = \{ \mathbf{S}(t) \mid t \in [t_0, t_f] \}
\]
where $t_0$ is the start time of behavior and $t_f$ is the end time of behavior. Different from events, behaviors only capture the self-motion characteristics of a trajectory, such as acceleration, deceleration, turning left, or moving straight, as illustrated in Fig. 1. Consequently, behaviors do not encompass the semantics of interactions between the trajectory and its surrounding environment. 

The status set \( \mathbf{S} \) represents a set of \( k \) statuses, where each status \( S_i \) corresponds to one scatterer at a specific moment.  These statuses capture the immediate semantic characteristics within a single snapshot, describing the self-distribution of power and delay, as well as the relative characteristics of multiple scatterers, respectively. This is formally represented as:
\[
\mathbf{S} = \{ S_1, S_2, \dots, S_k \}, \quad S_i \sim \mathcal{D}(\theta)
\]
where \( \mathcal{D} \) represents any potential distribution corresponding to a specific status (e.g., \( \mathcal{N} \) for normal distribution, \( \mathcal{P} \) for Poisson distribution, etc.), and \( \theta \) represents all relevant parameters. Status semantics are used to describe the fundamental environment without incorporating time. For example, in Fig. 1, the contribution of different scatterers, such as trees, vehicles, pedestrians, and buildings, to the channel can be considered as their corresponding status semantics, which are specifically influenced by factors such as the type, number, size, and relative position of the scatterers. As the basic semantic units in the channel semantics model, continuous status semantics enrich the channel trajectory.

Given the above definitions, the corresponding ISAC channel impulse response of status semantics $\mathbf{S}$ can be expressed as

\begin{equation}
h(\tau)= \sum_{k=1}^{K} \sum_{l=1}^{L(k)} a_l\cdot e^{j \varphi_l} \cdot \delta\left(\tau-\tau_l\right)
\end{equation}
where $\delta(\cdot)$ is the Dirac delta function, $L(k)$ is the number of MPCs related to $S_k$, $K$ is the number of status sementics, $a_l$ and $\tau_l$ are the amplitude and delay of the $l$th MPCs respectively. $\varphi_l$ is the phase and is assumed to be uniformly distributed in the range of $[0,2\pi]$.

For time-varying ISAC channels, the behaviors change continuously over time, while MPCs remain time-varying and independent at each time. Specifically, due to dynamic environments, each behavior semantics has own lifetime that appear and disappear at any time instant, and the amplitude and delay are time-varying within lifetimes. The set of all behavior semantics that exist at time instant $t_i$ are $B_i$, which is shown as follows:
\begin{equation}
B_i=B_{1, i} \cup B_{2, i} \cup, \ldots, B_{j, i} \cup, \ldots, B_{i-1, i} \cup B_{i, i}
\end{equation}
where $B_{j,i}$ is the behavior semantics that first appear at time $t_j$ and still exist at time $t_i$. The number of MPCs related to $S_k$ in $B_{j,i}$ is $L(t_{j,i},k)$. And the number of all MPCs at time $t_i$ is expressed as
\begin{equation}
L\left(t_i,k\right)=L\left(t_{i, i},k\right)+\sum_{j=1}^{i-1} L\left(t_{j, i},k\right)(0<j<i)
\end{equation}

For event semantics, the channel impulse response at time $t$ can be reconstructed based on the collection of MPCs in all sets of behavior semantics and status semantics, which can be thus expressed as

\begin{equation}
\begin{aligned}
h\left(t_i, \tau\right) = & \sum_{m=1}^{M} \sum_{k=1}^{K(m)} \left[ \sum_{j=1}^{i-1} \sum_{l=1}^{L(t_{j,i}, k)} a_l(t_i) e^{j \varphi_l} \delta\left(\tau - \tau_l(t_i)\right) \right. \\
& \left. + \sum_{l=1}^{L(t_i,k)} a_l(t_i) e^{j \varphi_l} \delta\left(\tau - \tau_l(t_i)\right) \right]
\end{aligned}
\end{equation}
where $K(m)$ is the number of MPCs related to status semantics $S_k$ in behavior semantics $B_m$, $M$ is the number of behaviors sementics. In this case, the $M$ behavior semantics and $K$ status semantics together consist of the event semantics.

\section{ISAC Channel Measurements}

\subsection{Measurement System}

\begin{figure*}[t]
    \centering
        \includegraphics[width=1\linewidth]{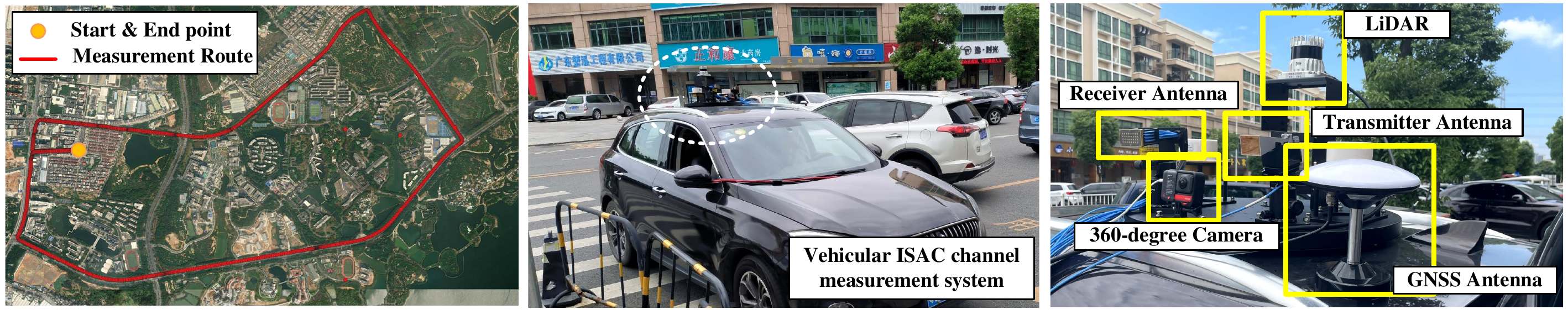}
    \caption{The ISAC channel measurement campaign.}
    \label{fig1}
\end{figure*}

The vehicular channel measurement system for ISAC scenarios is designed with key components, as shown in Fig. 2. These include a signal generator-based transmitter (TX), a signal digitizer-based receiver (RX), a 360-degree camera, a LiDAR, and a vehicle battery-powered supply. This setup enables simultaneous sensing of the electromagnetic and physical environments for comprehensive situational awareness. The TX employs the NI PXIe-5745 as a signal generator, providing 1 GHz baseband signals, which are up-converted to 28 GHz, amplified by a 28 dB power amplifier, and transmitted through horn antennas with a 15° beamwidth and 20 dB gain. The RX captures signals via 32-element array antennas with a 5 dB gain, amplified by low-noise amplifiers (LNAs), down-converted to baseband, and sampled by the NI PXIe-5745 signal analyzer. The TX and RX are co-located for mono-static ISAC channel measurement, synchronized using rubidium clocks and GNSS antennas for precise timing and positioning. For physical environment sensing, a 360-degree panoramic camera and LiDAR are positioned with the TX/RX, capturing RGB video, point-cloud data, and channel data simultaneously. This enables correlation of channel multipaths with environmental scatterers for semantic analysis of wave propagation. Power is supplied by an inverter converting the vehicle's 12V DC battery to 220V AC, with UPS ensuring stable and uninterrupted operation.

\subsection{Measurement Campaign}

The measurement campaign was conducted in the Songshanhu district of Dongguan, Guangdong, China, as shown in Fig. 3. The red line in Fig. 3(a) marks the vehicle trajectory for a single measurement route, covering a 15 km loop. The vehicle maintained an average speed of 30 km/h to ensure consistent data collection. TX and RX antennas were mounted on the vehicle roof, separated by 30 cm to avoid interference. A directional horn antenna at the TX focused the signal, and a 28 dBm power amplifier with a 20 dB gain antenna achieved a maximum two-way sounding distance of 50 meters, sufficient to capture most scatterers along the route. With a 1 GHz bandwidth, the system provides a 1 ns delay resolution, enabling the distinction of multipaths spaced as close as 0.3 meters. Fig. 4 illustrates the typical roadway environment, which includes elements like trees, buildings, streetlights, fences, greenbelts, and various vehicles. These elements are distributed across functional areas such as footpaths, flex zones, and travel lanes, forming a complex environment rich in semantic information for ISAC channel modeling. Their regular distribution patterns can be analyzed using event, behavior, and status semantics. Focusing primarily on the left side of the vehicle due to street symmetry, the campaign collected real-time electromagnetic data (via channel measurements) and physical data (via environmental sensing). This enables effective correlation of dynamic environmental changes with ISAC channel variations, providing insights into the interaction between the environment and electromagnetic wave propagation.

\section{Semantic Characterization and Modeling}

This section details the semantic characterization and modeling based on measurement data. First, semantic clustering is conducted using camera data. Subsequently, event, behavior, and status semantics are characterized based on the clustering results, and corresponding semantic models are developed.

\subsection{Semantic clustering on ISAC channel}

Extensive channel measurements show that multipath components (MPCs) are typically clustered, with MPCs in the same cluster often originating from the same scatterer \cite{A4,C2}. By simultaneously recording channel and visual data, and applying CV techniques, accurate semantic labels are assigned to channel clusters. This subsection presents a Density-Based Spatial Clustering of Applications with Noise (DBSCAN) -based semantic clustering method to effectively group multipaths and associate each cluster with its semantic label.

Firstly, the RGB images recorded by the camera are processed into depth images using the \textit{Depth Anything V2} model \cite{Net1}, a deep learning-based depth estimation framework combining Transformer and CNN architectures. This model effectively captures global context and local details, predicting scene depth information from a single image. Simultaneously, the RGB images undergo semantic segmentation using the \textit{Detectron2} model \cite{Net2}, developed by Facebook AI Research (FAIR). \textit{Detectron2} integrates deep convolutional neural networks and feature pyramid networks (FPN) to extract semantic information from images effectively. As shown in Fig. 5, the original image, its depth map, and semantic segmentation map are presented for the same video frame. The depth map accurately captures the distance between objects and the camera, while the segmentation results clearly identify objects and regions in the scene, enabling effective semantic information extraction.

\begin{figure}[t]
    \centering
        \includegraphics[width=1\linewidth]{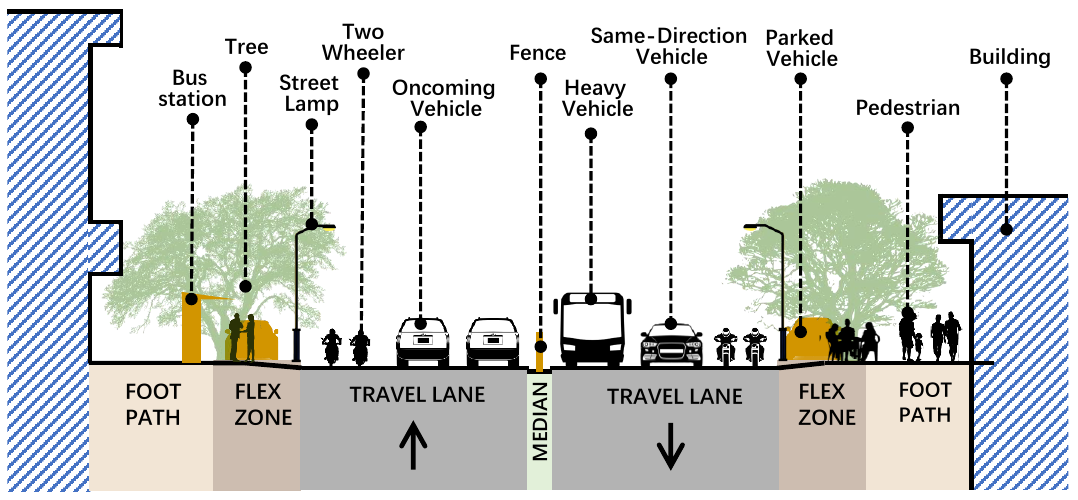}
    \caption{The layout of potential scatterers in measured environments.}
    \label{fig1}
\end{figure}

\begin{figure}[t]
    \centering
        \includegraphics[width=0.95\linewidth]{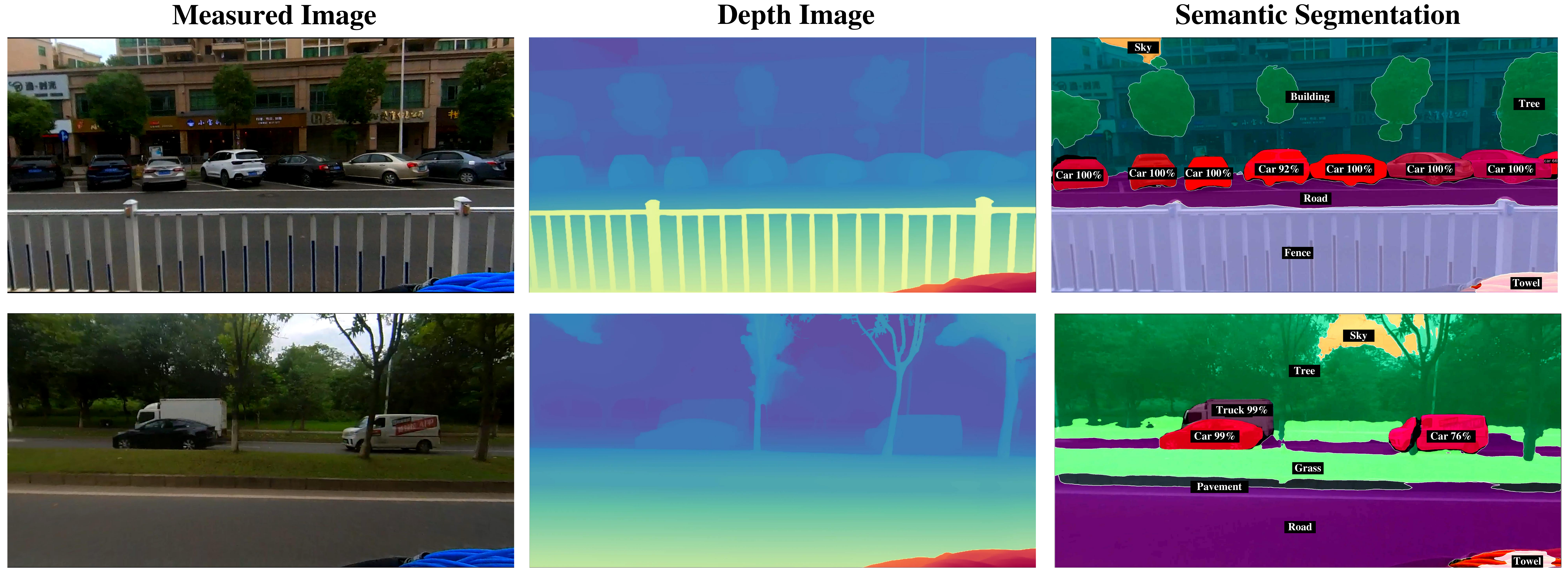}
    \caption{Semantic information extraction from measured data, which includes measured image, depth image, and semantic segmentation map.}
    \label{fig1}
\end{figure}

In this paper, depth value for each pixel in depth image is represented as \( D(x, y) \). The semantic label for each pixel in segmentation map is represented as \( S(x, y) \). The depth values are normalized to ensure that they are in the same physical scale as the propagation distances of the multipaths. The process is formalized as follows:
\begin{equation}
D'(x, y) = \frac{D(x, y) - \min(D)}{\max(D) - \min(D)} \cdot d_{\text{max}},
\end{equation}
where \( d_{\text{max}} \) is the maximum reachable physical distance in the channel measurement, which is set as 50 m, and \( D'(x, y) \) is the normalized depth value. Afterwards, the depth value for each semantic category is calculated as the propagation distance between the scatterers and the receiver. For each semantic category \( i \), the corresponding set of depth values is extracted as follows:
\begin{equation}
D_{i} = \{ D'(x, y) \mid (x, y) \in S(x, y), S(x, y) = i \}.
\end{equation}
The average depth distance for each semantic category is calculated as follows: 
\begin{equation}
 \bar{D}_{i} = \frac{\sum_{D' \in D_{i}} D'}{|D_{i}|}.
\end{equation}
where $|D_{i}|$ is the number of pixels in the semantic category $i$. This process results in a mapping between each semantic category and its corresponding average depth distance.

For channel modeling, power delay profiles (PDPs) are widely used to calculate the received multipaths with propagation delays, which is obtained by the square of the module of CIR as follows:
\begin{equation}
P (t, \tau)=|h(t, \tau)|^2
\end{equation}
where $h(t, \tau)$ is the CIR at time $t$ and delay $\tau$. The 6 dB threshold is used to filter out the background noise in the measurements, retaining the significant multipath components, as shown in Fig.6 (a). Subsequently, the multipaths in the temporal PDP are clustered based on the DBSCAN algorithm. Specifically, for a given multipath \( p \), its neighborhood is defined as the set of multipaths within a radius \( \varepsilon \), denoted as \( \mathcal{N}_{\varepsilon}(p) \):
\begin{equation}
\mathcal{N}_{\varepsilon}(p) = \{ q \mid d(p, q) \leq \varepsilon \},
\end{equation}
where \( d(p, q) \) represents the distance (e.g., delay difference or amplitude difference) between multipaths \( p \) and \( q \). A multipath \( p \) is classified as a \textit{core point} if the size of its neighborhood satisfies:
\begin{equation}
|\mathcal{N}_{\varepsilon}(p)| \geq \text{MinPts}.
\end{equation}

\begin{figure}[tbp]
\centering
\subfigure[]{\includegraphics[width=1.62in]{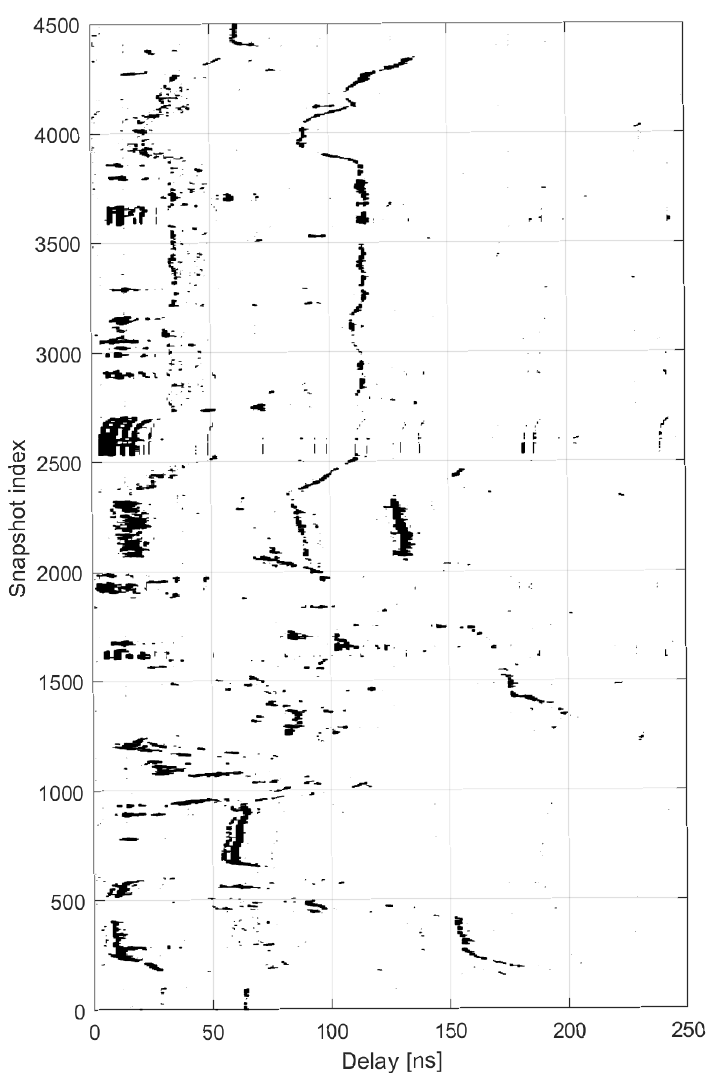}}
\subfigure[]{\includegraphics[width=1.83in]{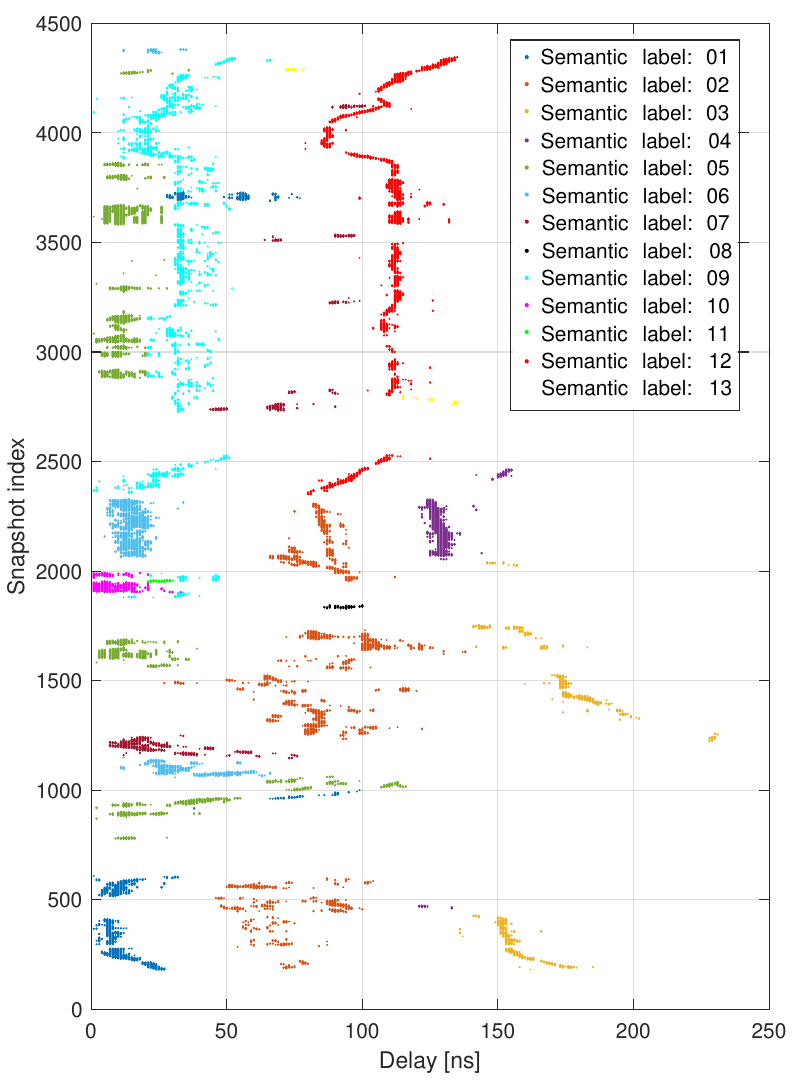}}
\subfigure[]{\includegraphics[width=3.5in]{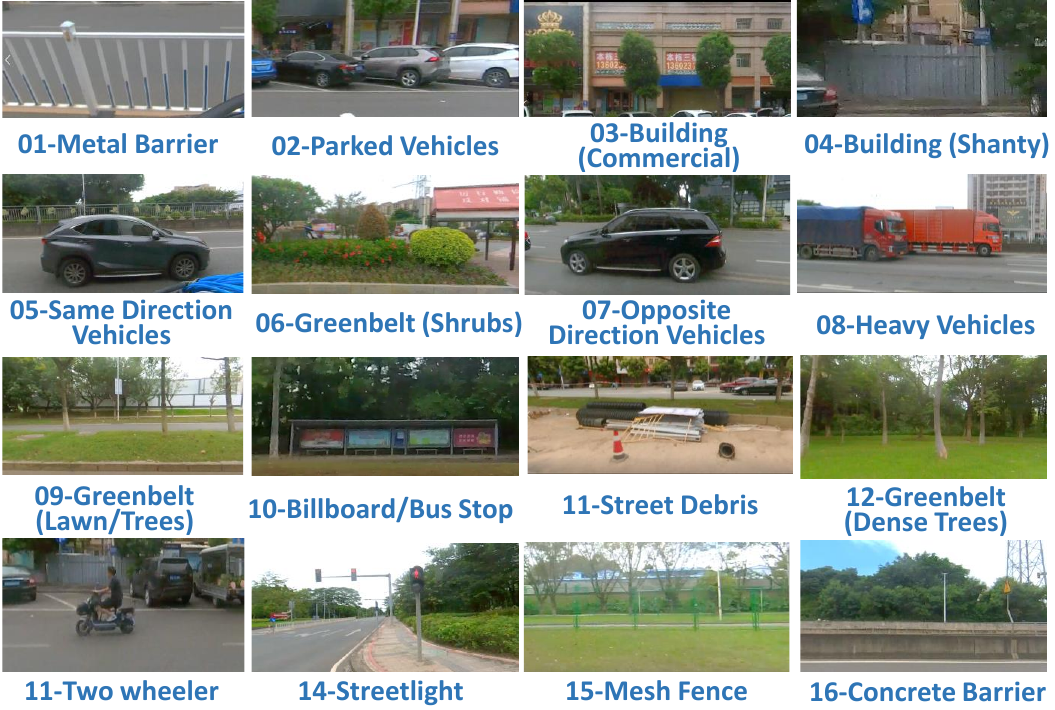}}
\caption{Semantic Clustering on ISAC Channel. (a) Multipath distribution patterns in the measured PDP. (b) Mapping channel clusters to semantic labels. (c) All semantic labels throughout the measurement.}
\label{fig}
\end{figure}

A multipath \( p \) is said to be \textit{density-reachable} from another multipath \( q \) if there exists a sequence of multipaths \( p_1, p_2, \dots, p_n \), where \( p_1 = q \), \( p_n = p \), and each multipath in the sequence lies within a distance \( \varepsilon \) of the next one. Additionally, the intermediate multipaths \( p_i \) must satisfy the condition of being core points, meaning their neighborhood contains at least \(\text{MinPts}\) multipaths. Two multipaths \( p \) and \( q \) are \textit{density-connected} if there exists a multipath \( o \) such that both \( p \) and \( q \) are density-reachable from \( o \). Based on these definitions, a cluster can be defined as the maximal set of multipaths that are mutually density-connected. Multipaths that do not satisfy these conditions or are not part of any cluster are considered noise. By applying the DBSCAN algorithm to cluster the multipaths, the resulting cluster set is denoted as $\{ C_1, C_2, \dots, C_k \}$. Similarly, to associate with semantic labels, we calculate the propagation distance for each cluster based on the delay. Since a mono-static ISAC sensing is used in the measurement, most of the multipaths are considered as single-hop. For each cluster \( C_k \), compute the center delay \( \bar{\tau_k} \) as follows:
\begin{equation}
\bar{\tau_k} = \frac{1}{|C_k|} \sum_{\tau \in C_k} \tau.
\end{equation}
Then, the propagation distance \( d_k \) based on the center delay can be calculated as:
\begin{equation}
d_k = \frac{c \cdot \bar{\tau_k}}{2},
\end{equation}
where \( c \) is the speed of light. The calculated propagation distance is matched with the average depth values of each semantic category to bind each cluster with its corresponding semantic label. For each cluster \( C_k \), find the semantic category \( I \) that has the closest average depth distance to the propagation distance \( d_k \):
\begin{equation}
I = \arg\min_{i} \left| d_k - \bar{D}_{i} \right|.
\end{equation}

After semantic clustering on the ISAC channel, channel clusters and their semantic labels are mapped, as shown in Fig. 6(b). To enhance the accuracy of semantic labels, we conducted manual inspection using video and point cloud data. For instance, the semantic label 'cars' was further divided into 'same-direction cars' and 'opposite-direction cars,' and misclassified labels were corrected. As summarized in Fig. 6(c), the measurement includes 16\text{+}1 semantic categories: 16 main categories such as trees, fences, and cars, and an additional 'other' category for instances without a clear semantic label. Notably, due to the sparsity of the millimeter-wave channel, most multipaths are clearly associated with a semantic label, with 'other' representing a very small proportion.

\subsection{Status Semantic Modeling}

Based on semantic clustering, channel multipaths are characterized for each semantic category, including the number of multipaths, delays, and power. Due to varying physical properties of scatterers, the distribution of multipath numbers differs across semantic clusters. Five common statistical models—Normal, Log-normal, Exponential, Gamma, and Weibull—are used to fit the multipath number distributions, with the best fit determined using the maximum likelihood method. Their mathematical expressions are as follows:

\begin{itemize}
  \item \textbf{Normal distribution:}
  \[
  f(x) = \frac{1}{\sqrt{2\pi\sigma^2}} \exp \left( -\frac{(x - \mu)^2}{2\sigma^2} \right)
  \]
  where \( \mu \) is the mean and \( \sigma^2 \) is the variance.

  \item \textbf{Log-normal distribution:}
  \[
  f(x) = \frac{1}{x\sigma \sqrt{2\pi}} \exp \left( -\frac{(\ln(x) - \mu)^2}{2\sigma^2} \right)
  \]
  where \( \mu \) is the mean and \( \sigma^2 \) is the variance of the natural logarithm of \( x \).

  \item \textbf{Exponential distribution:}
  \[
  f(x) = \lambda \exp(-\lambda x)
  \]
  where \( \lambda \) is the rate parameter.

  \item \textbf{Gamma distribution:}
  \[
  f(x) = \frac{\beta^\alpha}{\Gamma(\alpha)} x^{\alpha - 1} \exp(-\beta x)
  \]
  where \( \alpha \) is the shape parameter, \( \beta \) is the rate parameter, and \( \Gamma(\alpha) \) is the gamma function.

  \item \textbf{Weibull distribution:}
  \[
  f(x) = \frac{\alpha}{\lambda} \left( \frac{x}{\lambda} \right)^{\alpha - 1} \exp \left( -\left( \frac{x}{\lambda} \right)^\alpha \right)
  \]
  where \( \alpha \) is the shape parameter and \( \lambda \) is the scale parameter.
\end{itemize}

Fig. 7 shows the fitting results of the number of multipaths under different status semantics. Due to space limitations, only the fitting results for semantic labels 1–4 are presented here. The optimal distribution model varies across semantic labels: Gamma for Metal Barrier, Log-normal for Parking Vehicles and Building (Commercial), and Weibull for Building (Shanty). Table 2 summarizes the results for all semantic labels, showing that most distributions are Log-normal or Weibull, each covering six labels. The Log-normal distribution captures asymmetry and heavy-tailed characteristics with significant variability, while the Weibull distribution suits scenarios with consistent multipath structures. No semantic labels fit the Exponential distribution, as the multipath number exhibits more complex variation patterns beyond simple exponential decay.

\begin{figure}[t]
    \centering
        \includegraphics[width=1\linewidth]{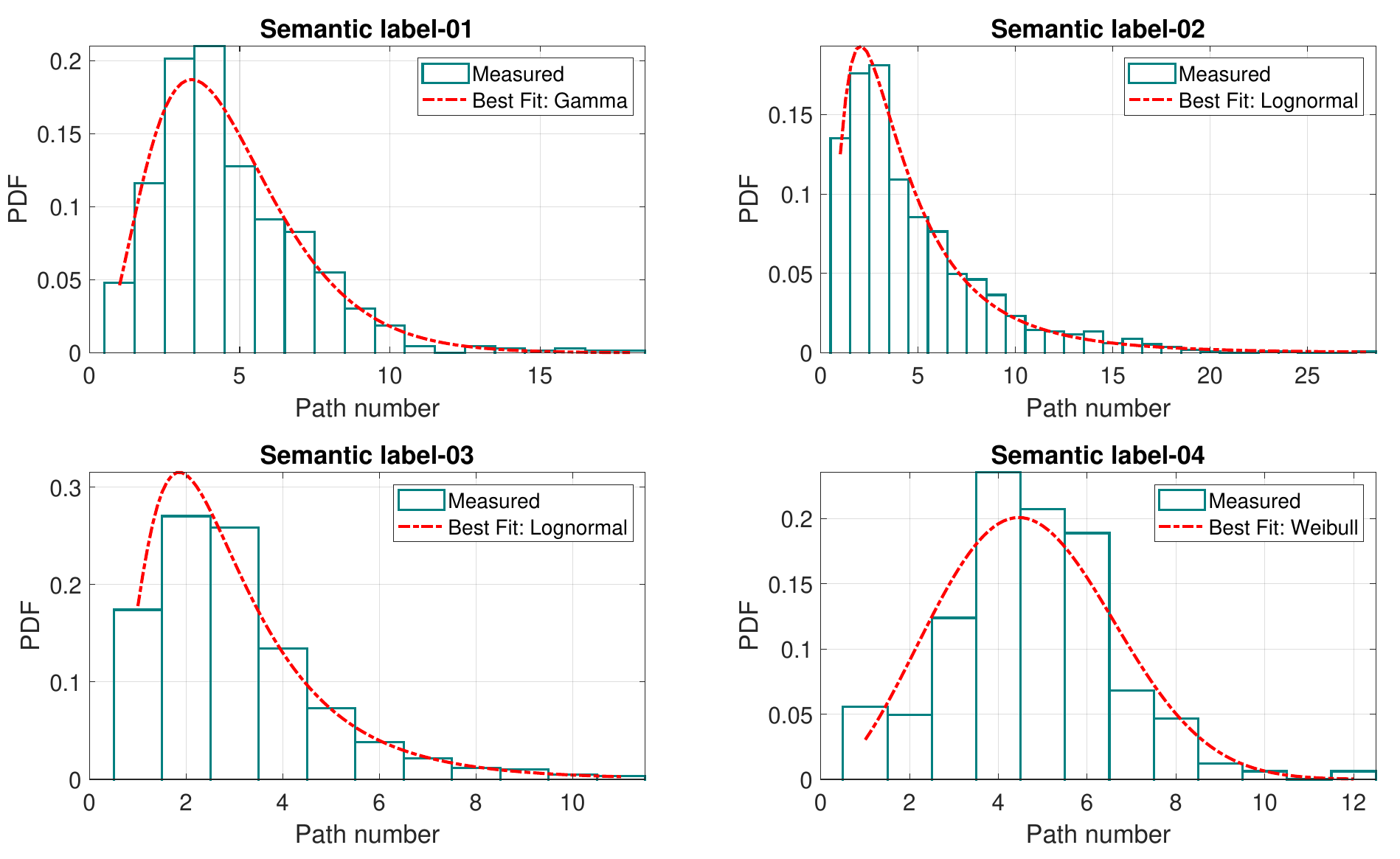}
    \caption{The fitting results of the number under different status semantics.}
    \label{fig1}
\end{figure}

In the proposed model, status semantics represent the inherent attributes of scatterers, independent of propagation distance. To ensure the characterization of each semantic cluster reflects only scatterer attributes, multipath power is first normalized using the free-space propagation loss model. The process is as follows:

\begin{equation}
p[dB] = p^*[dB] - 20 \log_{10}\left(\frac{\lambda}{4\pi \cdot \tau \cdot c}\right)
\end{equation}
where $\tau$ represents the delay of each multipath, $p$ and $p_i^*$ represent the normalized power and measured power of each multipath, respectively. To achieve customized semantic channel modeling, we describe the multipaths within each semantic cluster based on their relative power and delay with respect to the cluster centroid, instead of using absolute power and delay. In this way, when the cluster centroids change under different behaviors or events, the corresponding multipath distribution in the channel can be generated. We use the strongest multipath within the cluster as the cluster centroid, and the relative power and delay with respect to the cluster centroid are calculated as follows:
\begin{equation}
\Delta \tau = \tau_m-\tau , \Delta p = p_m - p
\end{equation}

We characterize the relative delay and relative power distributions of multipaths within semantic clusters. For relative delay, the physical properties of scatterers result in asymmetric distributions around the cluster centroid delay. Some semantic labels exhibit a longer tail on the later side (larger delays), while others have a longer lead on the earlier side (smaller delays). Overall, multipath density decreases as the distance from the cluster centroid increases. To capture this, exponential distributions are used to separately fit the relative delay distributions on the earlier and later sides of the cluster centroid. Fig. 8 presents the fitting results for semantic labels 1–4, demonstrating that even within the same semantic label, the relative delay distributions differ on the earlier and later sides, necessitating separate modeling. The exponential distribution provides a good fit, aligning with the natural decrease in multipath density as delay increases.

For the relative power, we model the relative power as a single-slope function of relative delay with parameters $\alpha $ and $\beta $, which is shown as follows:
\begin{equation}
\Delta p_i = \alpha \cdot \Delta \tau_i+\beta
\end{equation}
where $\Delta p_i$ and $\Delta \tau_i$ represent the relative power and delay of the $ i $-th multipath within the semantic cluster. Actually, the relationship between power and delay is not a deterministic function, thus,we also model the random residual power as a $ t $-location-scale distribution. Compared to the number of multipaths, which requires considering significant distribution differences across different status semantics, the characteristics of residual power are consistent across different semantics. Generally, the distribution of residual power typically exhibits heavy-tailed characteristics, reflecting the significant deviation caused by occasional strong reflections or scattering effects, and in practical channels, residual power may take negative values due to phase variations of the multipaths. The $ t$-location-scale distribution not only effectively models this heavy-tailed behavior but also handles negative values. Therefore, we directly use the $t$-location-scale distribution to fit the residual power distribution, which is shown as follows:

\begin{equation}
f(x; \mu, \sigma, \nu) = \frac{\Gamma\left(\frac{\nu+1}{2}\right)}{\sqrt{\nu\pi}\sigma\Gamma\left(\frac{\nu}{2}\right)} 
\left[1 + \frac{1}{\nu}\left(\frac{x-\mu}{\sigma}\right)^2\right]^{-\frac{\nu+1}{2}}
\label{eq:t_location_scale}
\end{equation}
where $\mu$ is the location parameter, $\sigma$ is the scale parameter, $\nu$ is the degrees of freedom, and $\Gamma(\cdot)$ represents the gamma function. Fig. 9 and 10 present the fitting results of the power decay and the residual power. Similarly, only the results for semantic labels 1–4 are shown due to space limitations. It can be observed that the power decay factor exhibits an overall negative linear relationship, and the residual power aligns well with the $ t $-location-scale distribution. Table 2 summarizes the modeling results for all status semantics. Based on this library, for any given cluster centroid, the corresponding multipaths under the semantic label can be generated, thereby forming the channel impulse response for a single snapshot.

\begin{figure}[t]
    \centering
        \includegraphics[width=1\linewidth]{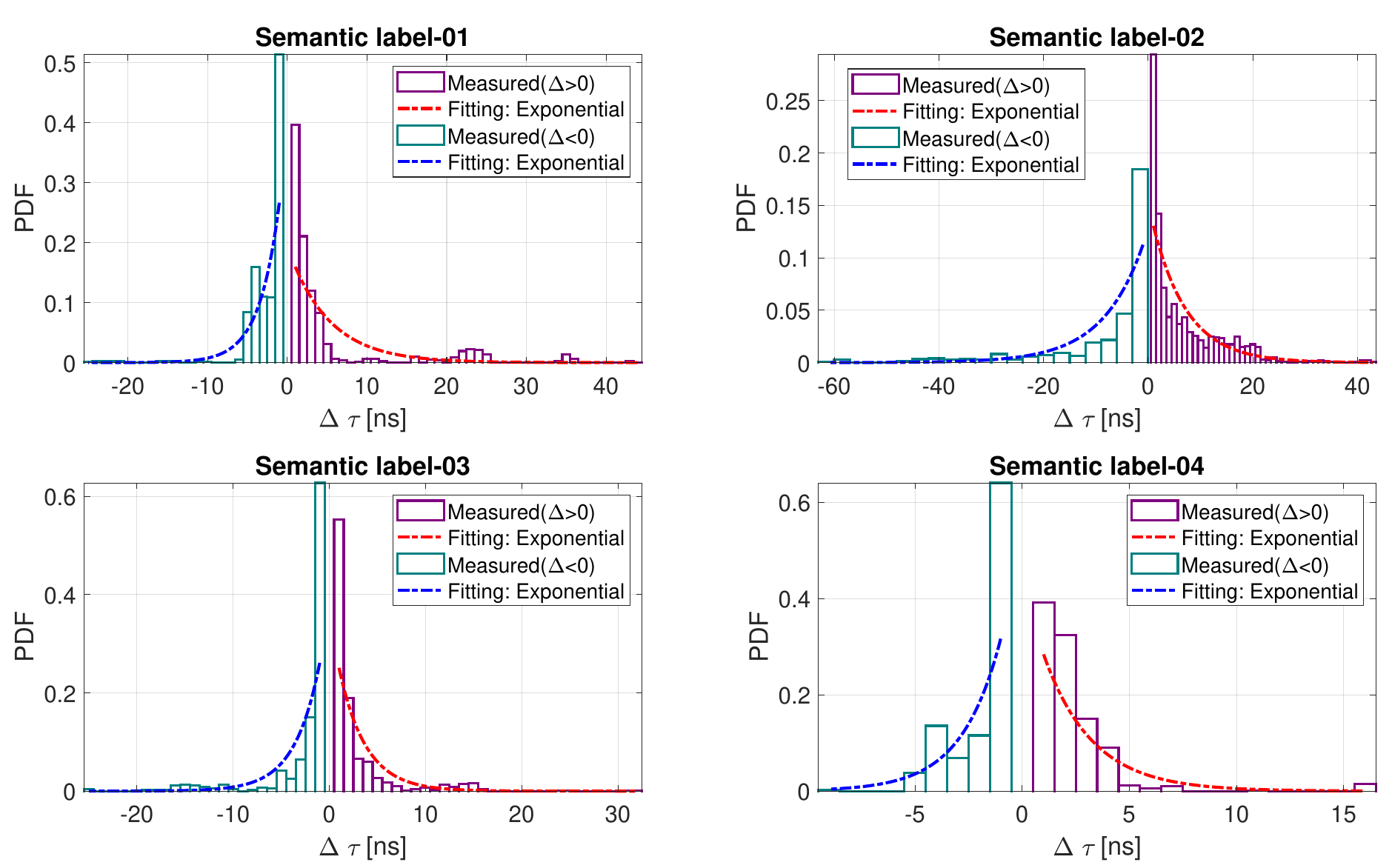}
    \caption{The fitting results of the delay under different status semantics.}
    \label{fig1}
\end{figure}

\begin{figure}[t]
    \centering
        \includegraphics[width=1\linewidth]{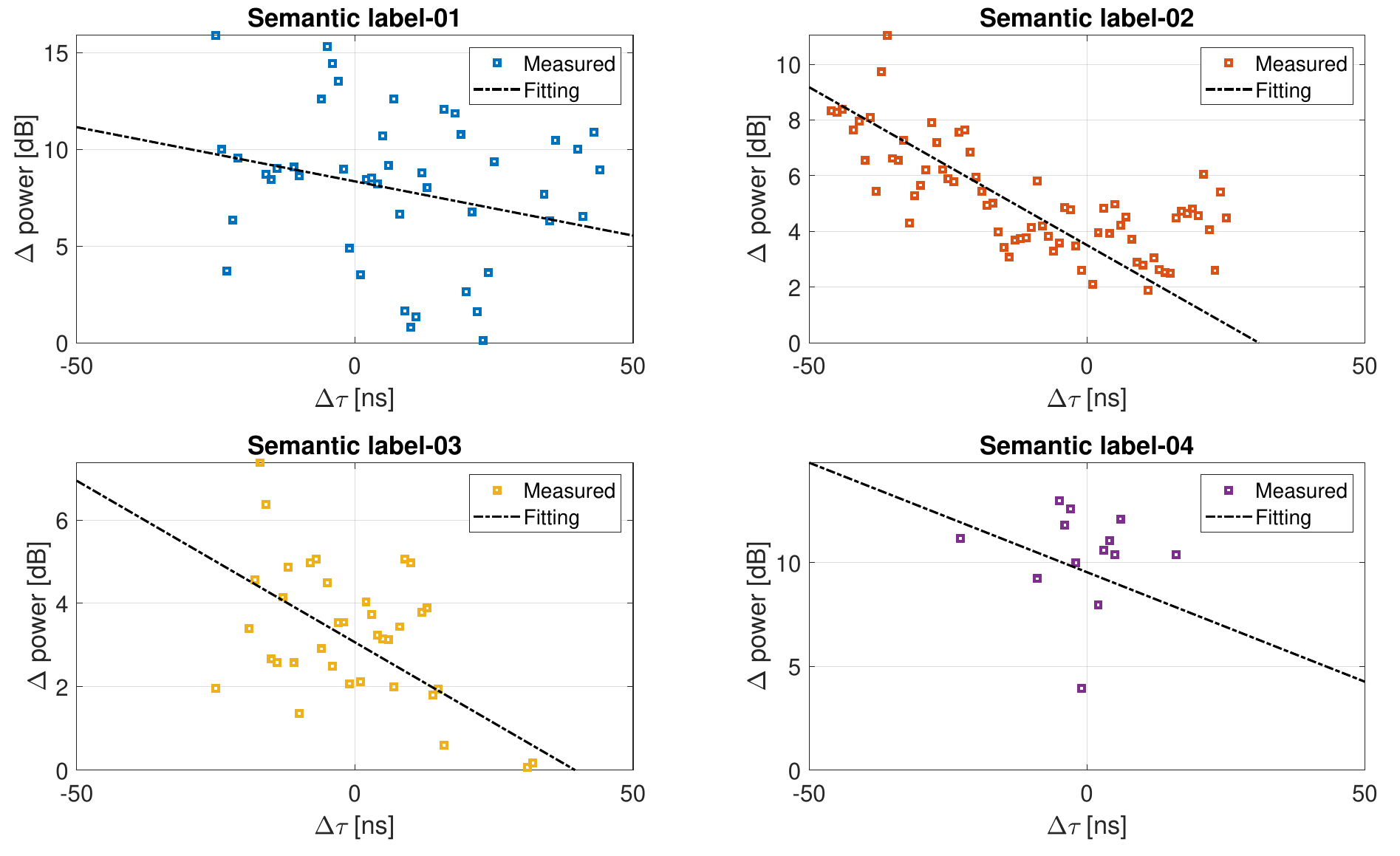}
    \caption{The fitting results of the power decay under different status semantics.}
    \label{fig1}
\end{figure}

\begin{figure}[t]
    \centering
        \includegraphics[width=1\linewidth]{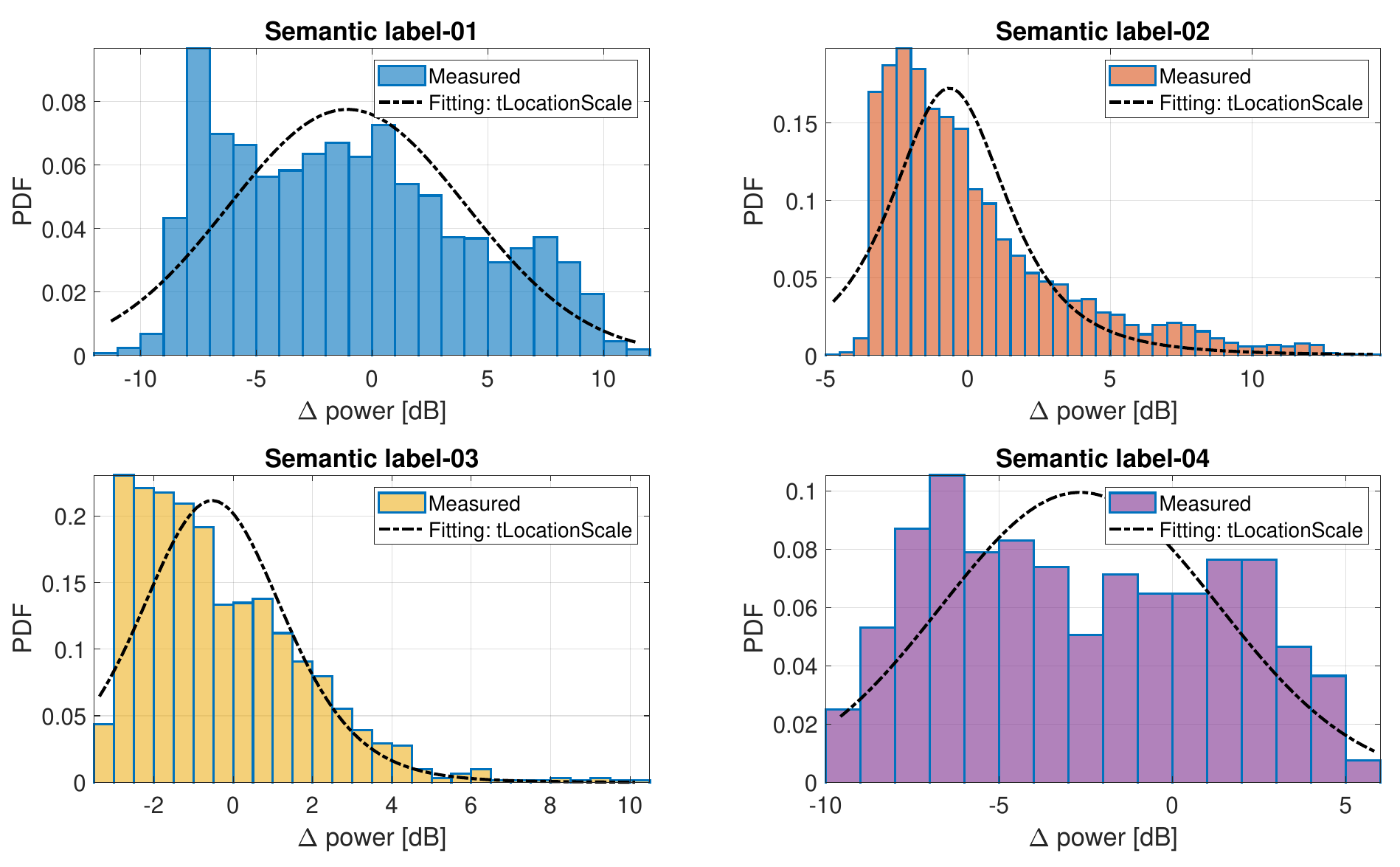}
    \caption{The fitting results of the residual power component under different status semantics.}
    \label{fig1}
\end{figure}


\begin{table*}[]
\centering
\caption{ The modeling library for status semantics.} \label{Table1Label}
\scalebox{0.9}{
\begin{tabular}{ccccccc}
\toprule
\multirow{2}{*}{\textbf{Status semantics}} & \multicolumn{2}{c}{\textbf{Multipath number}} & \multicolumn{2}{c}{\textbf{Multipath delay}}     & \multicolumn{2}{c}{\textbf{Multipath power}}                      \\
                                           & Distribution               & Parameters       & $\Delta \tau \textgreater{}0$ & $\Delta \tau\textless{}0$ & Power decay ($\alpha, \beta$) & Residual component ($\mu, \sigma,\nu$) \\ \hline
Metal Barrier (01)                         & Gamma ($\alpha, \beta$)      & (3.7022, 1.2574) & 5.1567                   & 2.5245                & (-0.0561, 8.3474)           & (-1.0574, 5.1470, 7.83e6)           \\
Parked Vehicles (02)                       & Log-normal ($\mu, \sigma$)  & (1.2876, 0.7601) & 6.5815                   & 7.8226                & (-0.1134, 3.5047)           & (-0.6478, 2.1218, 2.8688)           \\
Building-Commercial (03)                   & Log-normal ($\mu, \sigma$)  & (0.9509, 0.5773) & 2.7799                   & 2.611                 & (-0.0775, 3.0660)           & (-0.5287, 1.8278, 8.1347)           \\
Building-Shanty (04)                       & Weibull ($\alpha, \lambda$)   & (5.3113, 2.6741) & 2.2646                   & 1.8289                & (-0.1057, 9.5430)             & (-2.6583, 4.0122, 9.52e6)           \\
Same Direction Vehicles (05)               & Weibull ($\alpha, \lambda$)   & (11.4508,1.8621) & 10.5704                  & 5.8476                & (-0.1575, 11.0896)          & (-0.0768, 6.1096, 1.03e7)           \\
Greenbelt-Shrubs (06)                      & Weibull ($\alpha, \lambda$)   & (8.8260, 2.0461) & 5.3976                   & 4.2708                & (-0.1776, 3.9232)           & (-0.4330, 2.3026, 2.50e6)           \\
Opposite Direction Vehicles (07)           & Gamma ($\alpha, \beta$)      & (2.1789, 2.5215) & 5.9366                   & 4.8713                & (-0.1272, 4.9839)           & (-1.0940, 2.9148, 4.7956)           \\
Heavy Vehicles (08)                        & Weibull ($\alpha, \lambda$)   & (7.1248, 1.8978) & 2.4801                   & 5.2727                & (-0.0872, 1.8987)           & (-0.0282, 1.1254, 6.54e6)           \\
Greenbelt-Lawn/Trees (09)                  & Log-normal ($\mu, \sigma$)  & (1.0912, 0.5414) & 4.224                    & 3.9761                & (-0.2150, 2.7138)           & (-0.2693, 1.7752, 10.204)           \\
Billboard/Bus Stop (10)                    & Weibull ($\alpha, \lambda$)   & (8.4817, 1.3294) & 6.7883                   & 3.023                 & (-0.0426, 9.2137)           & (1.7503, 5.9625, 7.76e6)            \\
Street Debris (11)                         & Normal ($\mu, \sigma$)      & (6.1818, 2.7863) & 2.2857                   & 3.4186                & (-0.2987, 1.3597)           & (0.0209, 0.8566, 5.83e6)            \\
Greenbelt-Dense Trees (12)                 & Log-normal ($\mu, \sigma$)  & (1.2870, 0.6756) & 5.0038                   & 4.2969                & (-0.2402, 2.6727)           & (-0.0091, 1.8551, 6.6132)           \\
Two wheeler (13)                           & Log-normal ($\mu, \sigma$)  & (0.5883, 0.5298) & 2.5556                   & 1.5294                & (-0.0661, 1.4361)           & (-0.0711, 0.9983, 5.59e6)           \\
Streetlight (14)                           & Normal ($\mu, \sigma$)      & (7.9125, 4.0634) & 5.9459                   & 4.431                 & (-0.2448, 3.2563)           & (-0.2996, 1.9087, 1.95e6)           \\
Mesh Fence (15)                            & Gamma ($\alpha, \beta$)      & (6.3516, 0.5057) & 1.9422                   & 2.0333                & (-0.1002, 3.2019)           & (-0.5308, 1.7715, 6.07e5)           \\
Concrete Barrier (16)                      & Gamma ($\alpha, \beta$)      & (8.0071, 0.3023) & 1.0625                   & 1.4813                & (-0.3127, 4.8527)           & (-1.8761, 2.1243, 6.55e6)           \\ \hline
\end{tabular}}
\end{table*}

\subsection{Behavior Semantic Modeling}

\begin{figure}[t]
    \centering
        \includegraphics[width=0.8\linewidth]{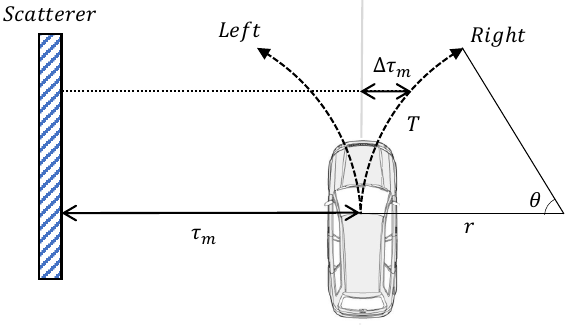}
    \caption{The deterministic relationship between the current cluster centroid’s delay $\tau_m$, turning angle $\theta$, behavior duration $T$, and the offset $\Delta \tau_m$.}
    \label{fig1}
\end{figure}

For different semantic clusters, behavioral semantics are modeled by the trajectory of the cluster centroid in the PDP, capturing variations in the centroid's delay and power. In the measurements, these behaviors are tied to the sensing terminal (the vehicle). Therefore, behaviors are classified into three categories: straight driving, left turning, and right turning. As the measurements focus on the left side of the vehicle, the impact of these behaviors on multipaths is primarily characterized as follows:

\begin{itemize}
    \item \textbf{Straight driving}: multipaths remain unchanged.  
    \item \textbf{Left turning}: multipath delays decrease.  
    \item \textbf{Right turning}: multipath delays increase.  
\end{itemize}

Furthermore, in actual channels, multipaths exhibit a birth-death process due to scattering dynamics. Consequently, multipaths can be modeled as four status: remaining unchanged, advancing, delaying, and transitioning between birth and death. We adopt a Markov chain to model the variations of multipaths. The statu transition matrix is defined as follows:  
\begin{equation}
\Pi = 
\begin{bmatrix}
\pi_{11} & \pi_{12} & \pi_{13} & \pi_{14} \\
\pi_{21} & \pi_{22} & \pi_{23} & \pi_{24} \\
\pi_{31} & \pi_{32} & \pi_{33} & \pi_{34} \\
\pi_{41} & \pi_{42} & \pi_{43} & \pi_{44} 
\end{bmatrix},
\end{equation}
where $\pi_{ij}$ denotes the probability of transitioning from statu $i$ to statu $j$. The status are defined as: 1) Unchanged: the multipath remains stable; 2) Advancing: The multipath delay decreases; 3) Delaying: The multipath delay increases; 4) Birth-death transition: the multipaths are either created or destroyed. These probabilities can be estimated through actual measurements, which is shown in Table 3. It can be observed that when the behavior semantics is straight driving, the multipath is more likely to remain unchanged. However, when the behavior semantics is left turning, the probability of advancing becomes higher. Besides, for all behaviors, the probabilities of path birth and death are approximately equal, each accounting for half. 

Although we can determine how multipaths change based on the statu transition matrix, obtaining the trajectory of the cluster centroid over time requires knowing the offset of delay. For the straight driving, we assume that each offset corresponds to a single delay bin. However, due to the determinism of the environment, the offset cannot be modeled statistically when turning. Typically, the turning radius and turning angle can uniquely determine the motion trajectory. Fig. 11 presents the relationship between the current cluster centroid’s delay $\tau_m$, andturning radius $r$, turning angle $\theta$, behavior duration $T$, and the offset $\Delta \tau_m$, which can be calculated as follows:
\begin{equation}
\Delta \tau_m = \frac{2\cdot r}{c}\cdot [1 - \cos\left( \frac{\theta}{T} \cdot \frac{1}{f_s} \right)]
\end{equation}
where $f_s$ is the channel sampling rate. The duration of different behavioral semantics can be fitted using a log-normal distribution, as shown in Fig. 12. It can be observed that the duration of straight driving is significantly longer than that of left and right turning. Additionally, since the sensing direction is on the left side in measurements, the distance to scatterers decreases during left turns, which promotes the survival of multipaths. Therefore, the duration of left turning is slightly longer than that of right turning.

\begin{table*}[]
\centering
\caption{The modeling library for behavior semantics.} \label{Table1Label}
\scalebox{0.9}{
\begin{tabular}{@{}cccccccc@{}}
\toprule
\textbf{Behavior semantics}                &             & Unchanged & Advancing & Delaying & Birth-death & Duration ($\mu$, $\sigma$) & Variation of power ($\mu$, $\sigma$) \\ \midrule
\multirow{4}{*}{\textbf{Straight driving}} & Unchanged   & 0.8851    & 0.0500    & 0.0445   & 0.0204      & \multirow{4}{*}{(4.2741, 1.6653)}                    & \multirow{4}{*}{(-0.0013, 0.9854)}                             \\
                                           & Advancing   & 0.8650    & 0.1033    & 0.0076   & 0.0240      &                                                      &                                                                \\
                                           & Delaying    & 0.8589    & 0.0087    & 0.1029   & 0.0295      &                                                      &                                                                \\
                                           & Birth-death & 0.4463    & 0.0326    & 0.0307   & 0.4904      &                                                      &                                                                \\ \midrule
\multirow{4}{*}{\textbf{Left turning}}     & Unchanged   & 0.7723    & 0.1546    & 0.0310   & 0.0421      & \multirow{4}{*}{(3.4918, 1.2863)}                    & \multirow{4}{*}{(0.0434, 0.9783)}                              \\
                                           & Advancing   & 0.8592    & 0.1173    & 0.0013   & 0.0212      &                                                      &                                                                \\
                                           & Delaying    & 0.9375    & 0.0156    & 0.0156   & 0.0313      &                                                      &                                                                \\
                                           & Birth-death & 0.4486    & 0.0324    & 0.0216   & 0.4973      &                                                      &                                                                \\ \midrule
\multirow{4}{*}{\textbf{Right turning}}    & Unchanged   & 0.7646    & 0.0292    & 0.1881   & 0.0181      & \multirow{4}{*}{(3.7115, 0.8170)}                    & \multirow{4}{*}{(-0.0376, 0.9711)}                             \\
                                           & Advancing   & 0.8268    & 0.0787    & 0.0157   & 0.0787      &                                                      &                                                                \\
                                           & Delaying    & 0.9204    & 0.0004    & 0.0610   & 0.0183      &                                                      &                                                                \\
                                           & Birth-death & 0.4641    & 0.0015    & 0.0537   & 0.4807      &                                                      &                                                                \\ \bottomrule
\end{tabular}}
\end{table*}

\begin{figure}[t]
    \centering
        \includegraphics[width=0.8\linewidth]{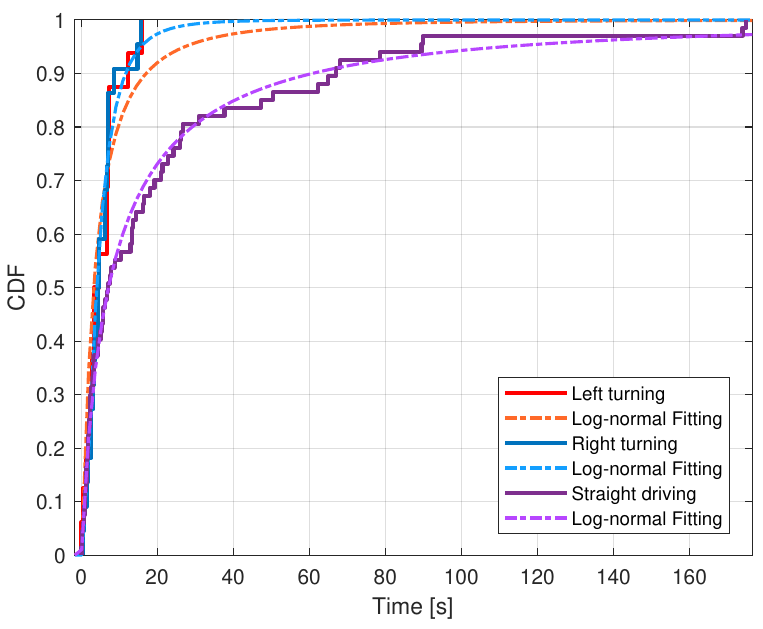}
    \caption{The fitting results of the duration under different behavior semantics.}
    \label{fig1}
\end{figure}

\begin{figure}[tbp]
\centering
\subfigure['straight driving']{\includegraphics[width=1.1in]{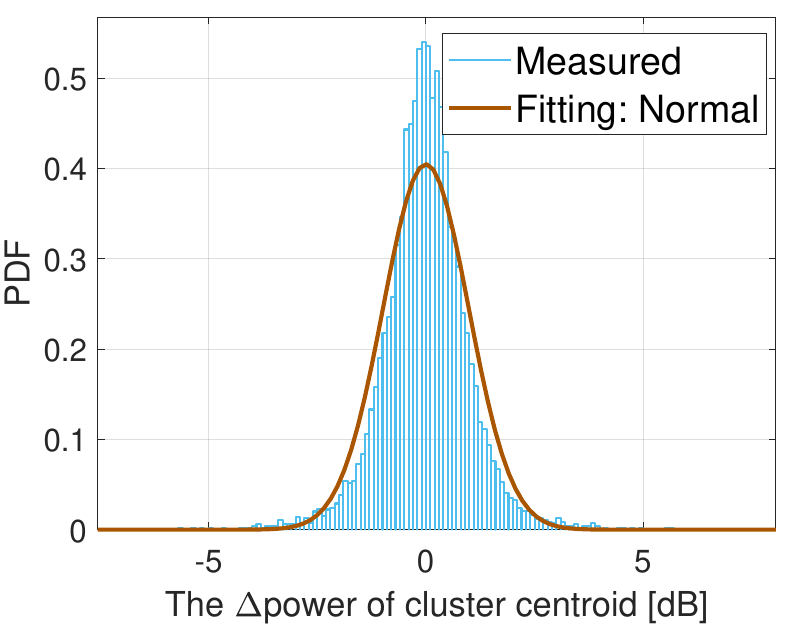}}
\subfigure['left turning']{\includegraphics[width=1.1in]{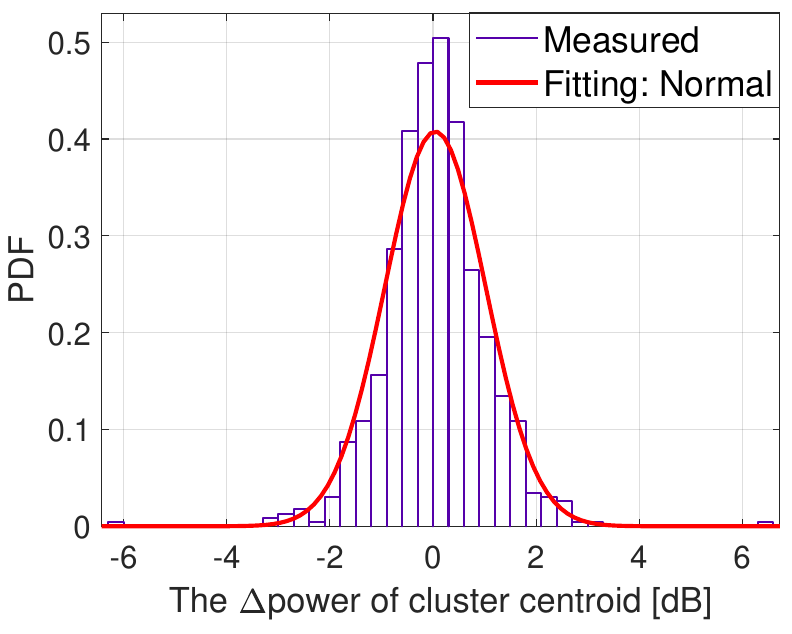}}
\subfigure['right turning']{\includegraphics[width=1.1in]{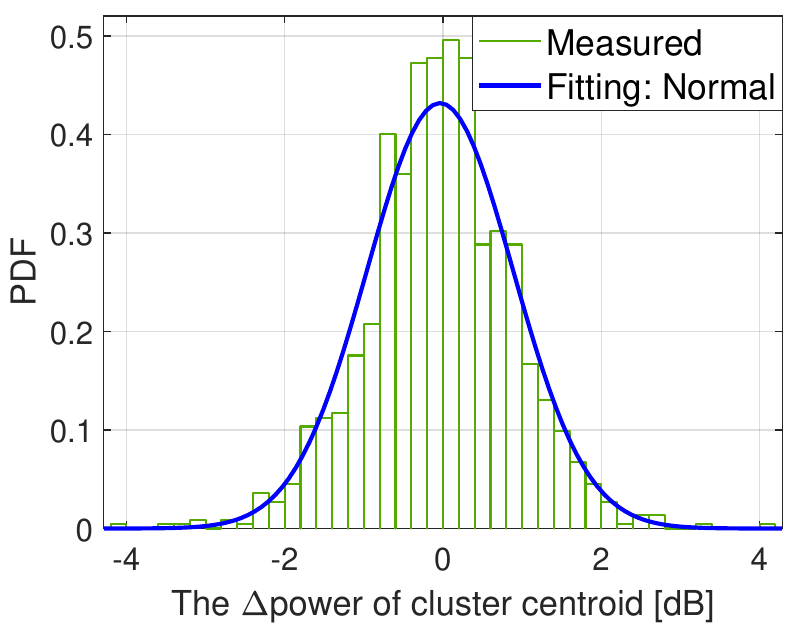}}
\caption{The fitting results of the variation in the cluster centroid's power under different behavior semantics.}
\label{fig}
\end{figure}

The variation in the cluster centroid's power can be modeled by fitting a normal distribution, as shown in Fig. 13. It can be observed that the power variation for the cluster centroid follows a normal distribution, with the mean centered around 0, and the proportions of increase and decrease being approximately balanced.

\subsection{Event Semantic Modeling}

For event semantic modeling, the relationships between status and behavior semantics are analyzed. Each cluster centroid is treated as a "word," each snapshot as a "sentence" of centroids, and a continuous channel as a "text" of multiple snapshots. Using a behavior correlation matrix and status co-occurrence matrix, the multipath topology for event semantics can be generated. Additionally, users can define custom semantic "texts" to create desired channels.

The behavior correlation matrix illustrates the relationship between behaviors and status, with each element representing the probability of dominant status semantics under a given behavior. Based on this matrix, we can determine how specific behaviors influence the appearance or disappearance of status semantics. Fig. 14 shows the behavior correlation matrix from the measurements. The most strongly correlated statuses across all behaviors are greenbelt lawn/trees (label 09) and greenbelt dense trees (label 12), reflecting their frequent presence during driving, consistent with typical road layouts. Status semantics are richer during "straight driving," indicating more diverse road conditions. For "left turning" and "right turning," the most correlated statuses are metal barrier (label 01) and same direction (label 05), respectively. This is due to the sensing direction being on the left: left turns bring the vehicle closer to the central median barrier, while right turns provide a wider view of vehicles moving in the same direction.

\begin{figure}[t]
    \centering
        \includegraphics[width=0.8\linewidth]{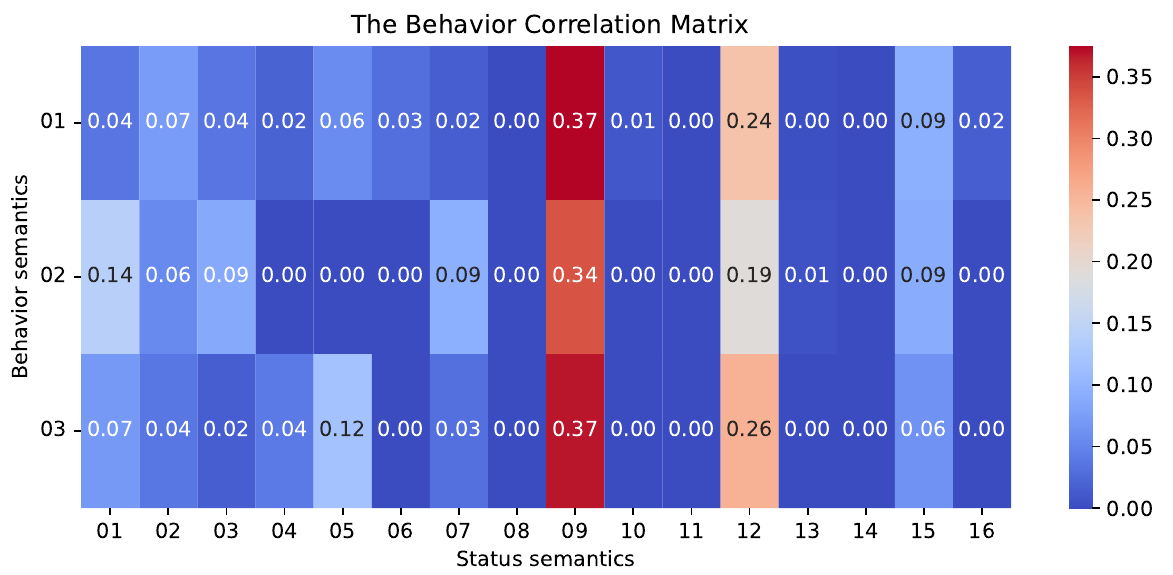}
    \caption{The behavior correlation matrix for the measurements.}
    \label{fig1}
\end{figure}

The status co-occurrence matrix captures the relationships between status semantics within the same snapshot, with each element representing the probability of one status coexisting with another. This matrix models the compatibility or exclusivity of different statuses. Fig. 15 shows the status co-occurrence matrix from the measurements, revealing three main relationships: independent, symmetric, and asymmetric. Independent relationships occur when a status exists only in snapshots without other statuses, e.g., concrete barrier (label 16). Symmetric relationships involve nearly equal co-occurrence probabilities, such as metal barrier (label 01) and parked vehicles (label 02), with probabilities of 0.30 and 0.39, respectively. Asymmetric relationships show significant differences in co-occurrence probabilities, e.g., greenbelt lawn/trees (label 09) has a 0.03 probability of co-occurring with street debris (label 11), while the reverse probability is 0.39. The status co-occurrence matrix quantifies these relationships, enabling the generation of realistic multipath topologies for channel semantic modeling.

Based on the multipath topology, we can identify the status semantics present at a specific moment. To determine the centroids of each semantic cluster, the initial delay and power of the centroids are also required. Fig. 16 illustrates the value ranges for delay and power of the centroids, showing significant variation across different status semantics. For instance, buildings correspond to larger delays, while metal barriers have smaller delays, reflecting the typical distribution of scatterers in real-world environments.

\begin{figure}[t]
    \centering
        \includegraphics[width=0.8\linewidth]{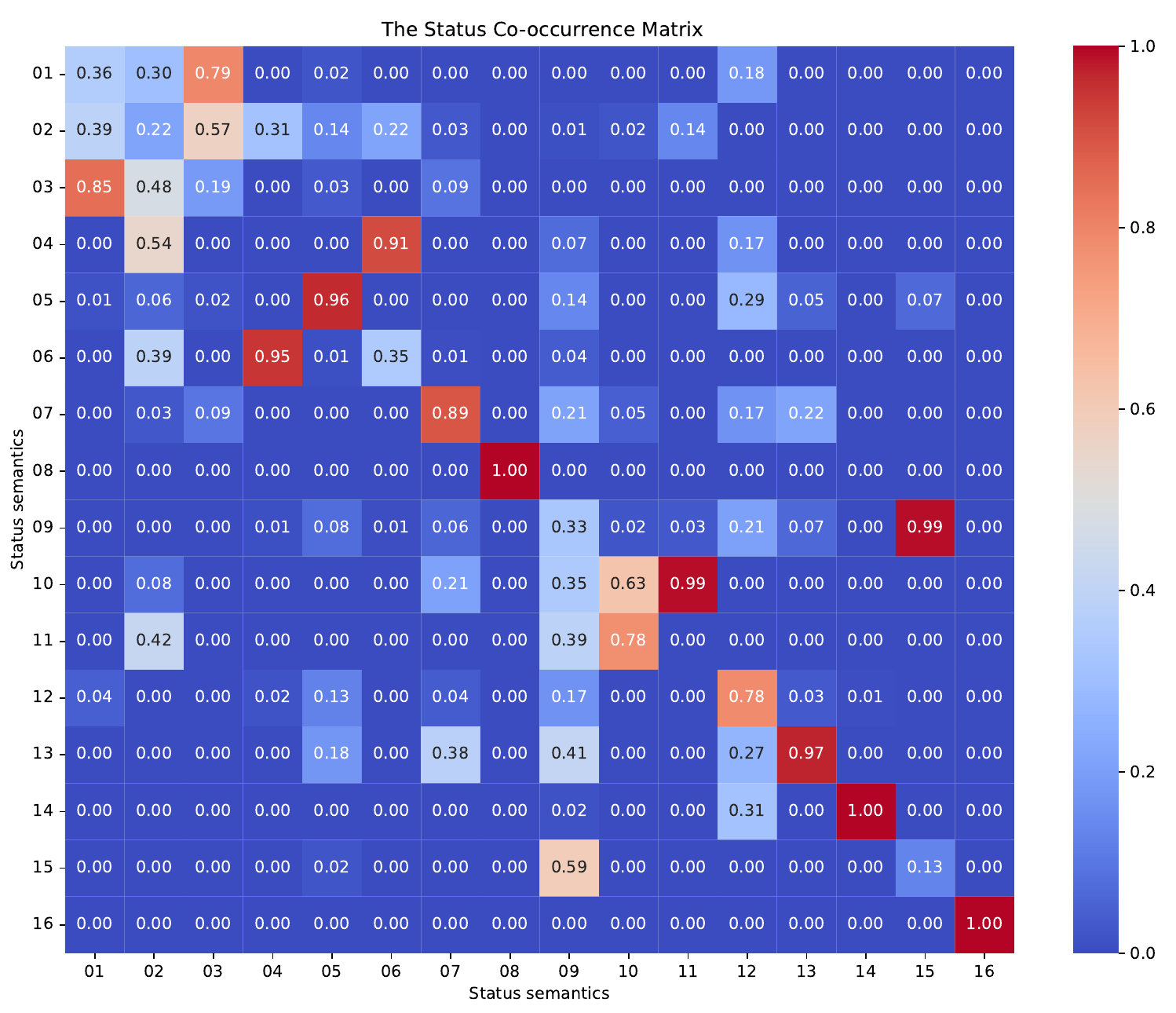}
    \caption{The status co-occurrence matrix for the measurements.}
    \label{fig1}
\end{figure}

\begin{figure}[tbp]
\centering
\subfigure{\includegraphics[width=1.4in]{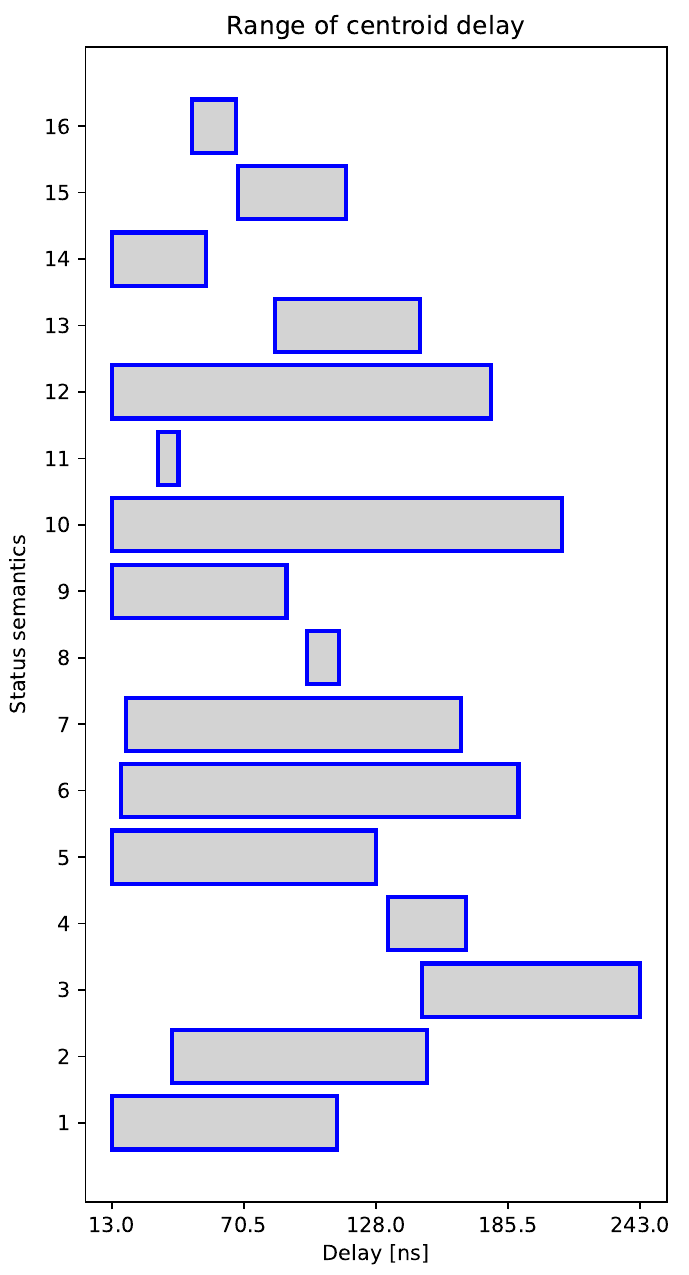}}
\subfigure{\includegraphics[width=1.4in]{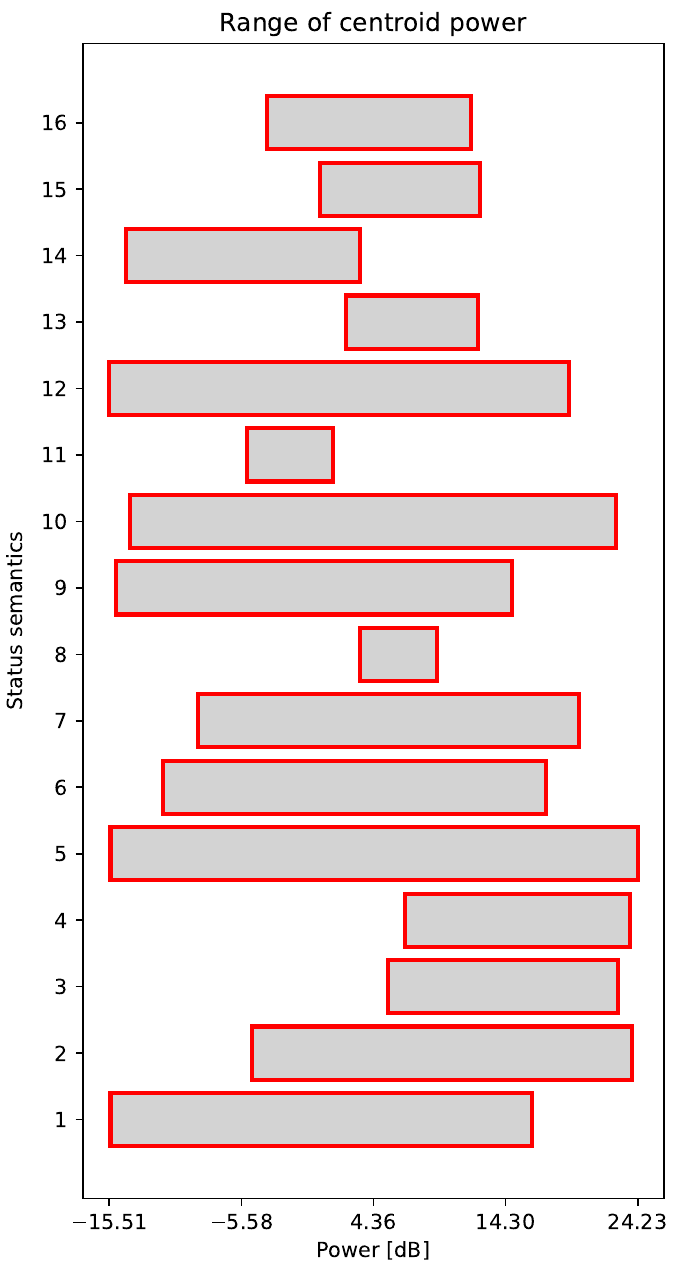}}
\caption{The ranges for the delay and power of the centroids.}
\label{fig}
\end{figure}

\section{Implementation and valiadation}

\begin{figure}[tbp]
\centering
\subfigure{\includegraphics[width=1.7in]{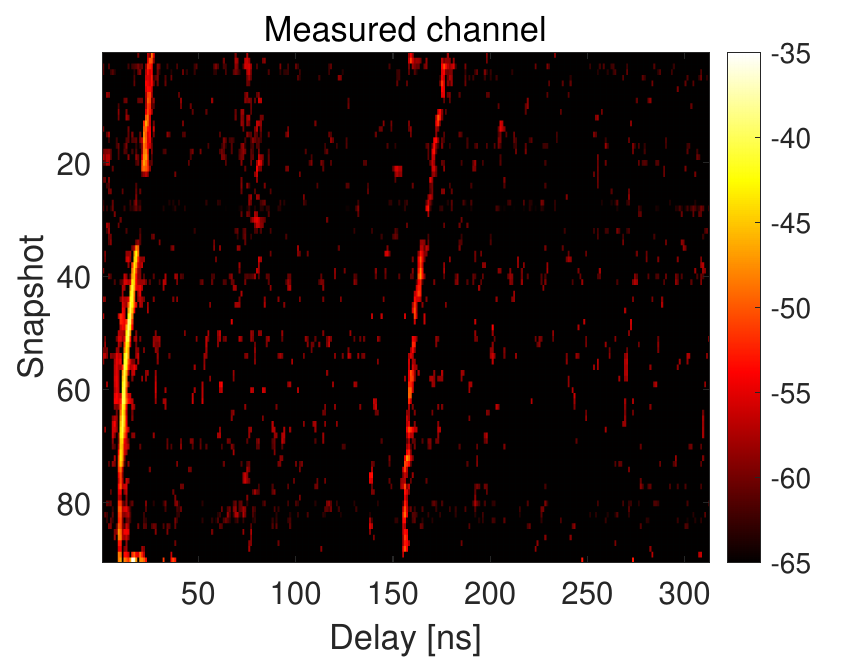}}
\subfigure{\includegraphics[width=1.7in]{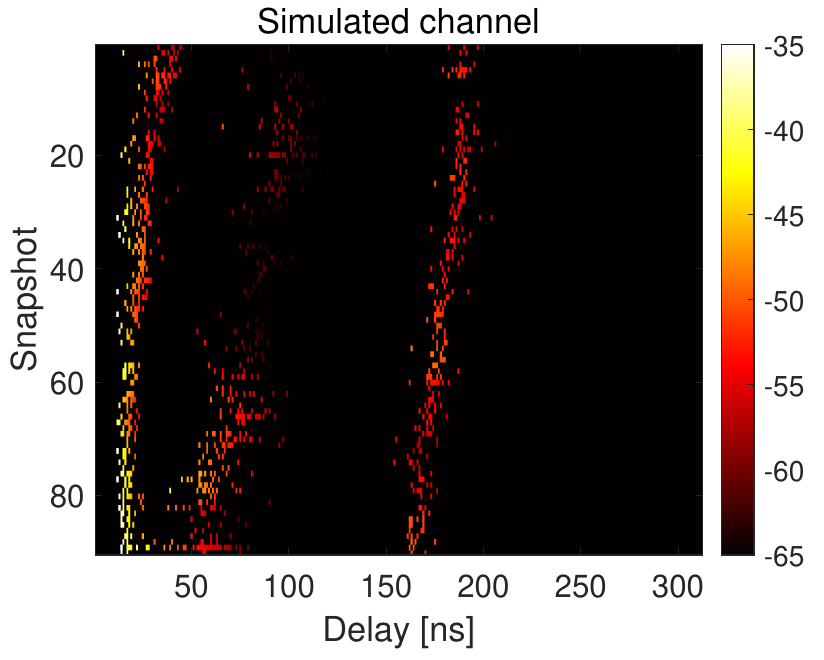}}
\caption{The measured and simulated PDPs under event semantics ``driving in an urban area and slowly turning left and merging into the center lane".}
\label{fig}
\end{figure}

\begin{figure}[t]
    \centering
        \includegraphics[width=0.8\linewidth]{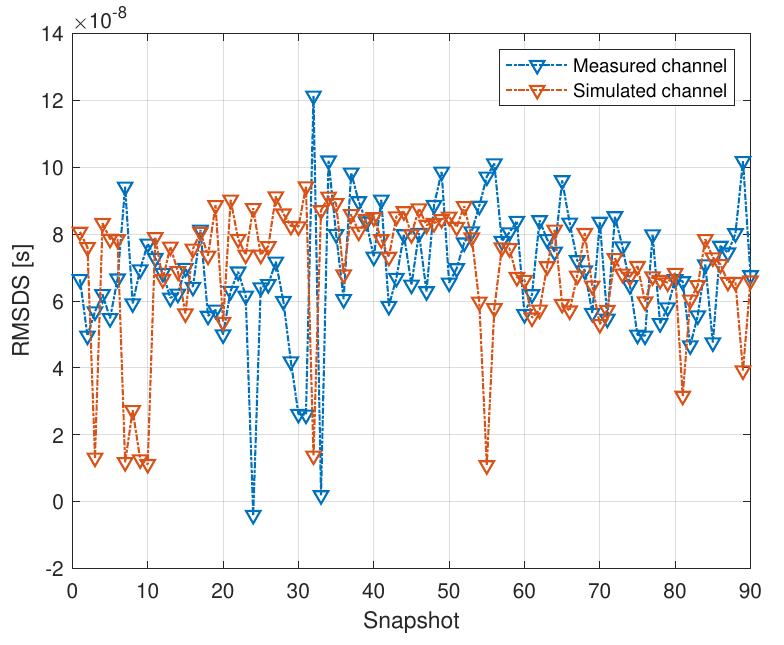}
    \caption{The comparison of RMSDS under event semantics "driving in an urban area and slowly turning left and merging into the center lane".}
    \label{fig1}
\end{figure}

\begin{figure}[t]
    \centering
        \includegraphics[width=0.8\linewidth]{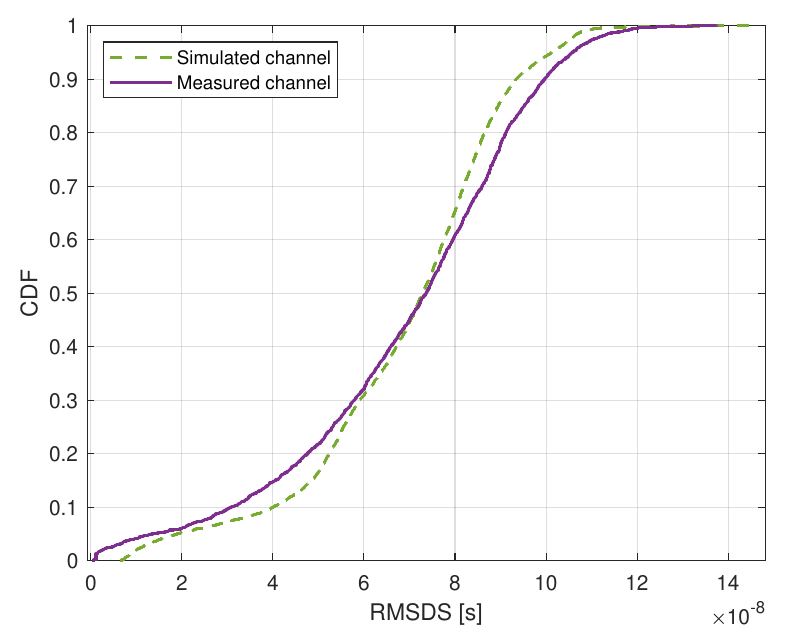}
    \caption{The global CDF between measured channel and simulated channel.}
    \label{fig1}
\end{figure}

\begin{figure}[tbp]
\centering
\subfigure[\centering Driving straight in a suburban area]{\includegraphics[width=3in]{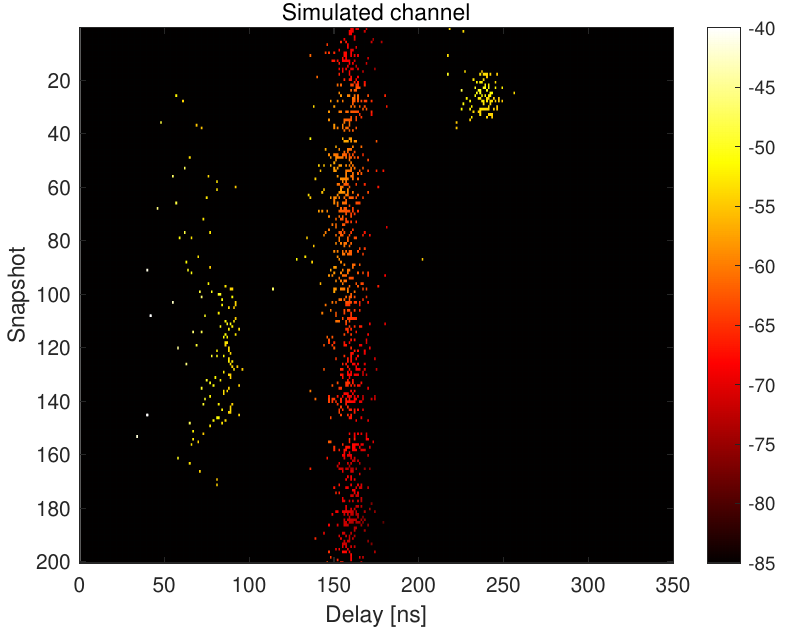}}
\subfigure[\centering Turning right at an intersection in an urban area]{\includegraphics[width=3in]{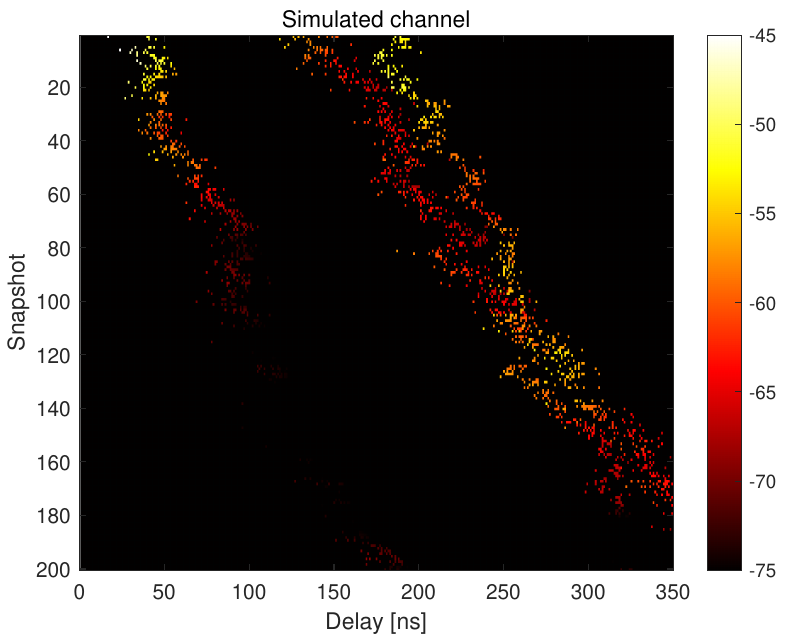}}
\caption{The simulated ISAC channel with customized semantics.}
\label{fig}
\end{figure}
Based on the modeling results for status semantics, behavior semantics, and event semantics in the previous section, we can generate channel models that represent the desired semantics. Specifically, the channel with semantics modeling process can be described as follows:

\begin{enumerate}
    \item \textbf{Input Initial Sequence:} Provide a predefined behavior sequence (e.g., [1, 2, 3, 3, 3, 2, 1,...]) as the input to the model. For each behavior using a normal distribution to generate realistic behavior duration distributions.
    
    \item \textbf{Event Map Modeling:} Determine the primary status semantics corresponding to each behavior semantic based on the \textit{behavior correlation matrix}, and identify coexisting status using the \textit{statu co-occurrence matrix}.
    
    \item \textbf{Centroid Initialization:} Based on the identified semantics, initialize the initial delay and initial power of semantic cluster centroids according to occurrences range fittings. 
    
    \item \textbf{Behavior Trajectory Modeling:} Based on the Markov chain model and delay/power variation models, simulate the temporal dynamics of centroids, including variations in delay and power, to reflect the time-varying characteristics of different semantic clusters.
    
    \item \textbf{Status Distribution Modeling:} Based on the statistical distribution model of the number of multipaths, delays, and power, simulate the multipath components within semantic clusters, ensuring that the resulting multipath topology aligns with the physical properties of the scatterers.
    
    \item \textbf{Output Channel Coefficient:} Generate channel coefficients based on the time-varying multipaths, which comprehensively describes the spatiotemporal distribution characteristics of the expected channel semantic representation.
\end{enumerate}

The final channel impulse response output represents high-level semantic information while accurately capturing the channel's physical and dynamic characteristics, making it suitable for multi-scenario simulation and performance evaluation. The model's performance is validated from two perspectives: a) accuracy in measured scenarios, and b) generalization across customized scenarios.

A segment of the measured route, described as "driving in an urban area, slowly turning left, and merging into the center lane," was selected to validate the accuracy of channel semantic representation in local details. Video recordings indicate this segment primarily involves metal barriers, parked vehicles, and commercial buildings, with the vehicle behavior being a left turn. Fig. 17 compares the measured and simulated time-varying PDPs, showing general consistency. Fig. 18 presents the corresponding root mean square delay spread (RMSDS), with average RMSE values of 67 ns for the measured channel and 69 ns for the simulated channel, demonstrating high accuracy.

For global channel characteristics, we simulated the overall channel for the entire measured scenario. Since the synthesized event semantics cannot fully replicate actual road conditions, the positions of simulated multipaths in the PDP lack direct comparability with the measured data. Thus, the focus was on second-order statistical parameters, as shown in Fig. 19. The results demonstrate that the channel model effectively represents propagation characteristics while capturing complex semantics over a large-scale route.

For the generalization validation across customized scenarios, we defined two events to demonstrate the generalization of channel semantic representation: 'driving straight in a suburban area' and 'Turning right at an intersection in an urban area'. Fig. 20 presents the simulated channel results under customized semantics. For the simulated channel semantics, in a suburban driving scenario, the predominant scatterers are long-term dense trees and short-term same-direction and opposite-direction vehicles. When turning right at an intersection, predominant scatterers such as the central median, greenbelt, and commercial buildings gradually move away. The generated channels align closely with expectations, exhibiting flexible and extensible scattering characteristics consistent with real-world environments while satisfying predefined semantic requirements and ensuring diversity.

\section{Conclusion}

With the evolution from 5G to 6G, conventional channel models, which primarily focus on physical properties, face significant challenges in representing the semantic information embedded in the environment, thereby limiting the performance evaluation of ISAC systems. This paper proposes a novel framework for ISAC channel modeling from a conceptual event perspective, which decomposes complex channel characteristics into composable and extensible multi-level semantic characterization, enabling flexible adjustments of channel models for different events without requiring a complete reset. Firstly, we characterize channel semantics as status semantics, behavior semantics, and event semantics, corresponding to channel multipaths, channel trajectories, and channel topology, respectively. By leveraging realistic channel measurement at 28 GHz, the proposed framework enables the extraction of semantic information through advanced algorithms such as depth estimation and semantic segmentation. The characterization of status semantics is achieved by fitting statistical distributions to multipath properties, while behavior semantics are captured using Markov chains to describe the dynamic variations of multipaths. Event semantics are represented using a co-occurrence matrix, associating dynamic behaviors with environmental statuses. The results validate the ability of the proposed model to not only accurately simulate channels but also incorporate rich, interpretable semantic information. Furthermore, the framework demonstrates generalization to support customized semantics, providing a flexible and extensible foundation for channel modeling and simulation, which can be used for ISAC intelligence system design and performance evaluation.

\bibliographystyle{IEEEtran}

\bibliography{refs.bib}

\end{document}